\documentclass[showkeys,showpacs,aps]{revtex4}

\usepackage{amsmath}
\usepackage{bm}
\usepackage{graphicx}
\usepackage{longtable}

\begin{document}

\title[Orthogonal Basis Set of Binocular Pupil]{Set of Orthogonal Basis Functions over the Binocular Pupil}

\author{Richard J. Mathar}
\homepage{http://www.strw.leidenuniv.nl/~mathar}
\email{mathar@strw.leidenuniv.nl}
\affiliation{Leiden Observatory, Leiden University, P.O. Box 9513, 2300 RA Leiden, The Netherlands}
\thanks{Supported by the NWO VICI grant 639.043.201 to A. Quirrenbach,
``Optical Interferometry: A new Method for Studies of Extrasolar Planets.''}


\date{\today}

\begin{abstract}
Sets of orthogonal basis functions over circular areas---often representing
pupils in optical applications---are known in the literature
for the full circle (Zernike or Jacobi polynomials) and the annulus.
Here, an orthogonal set is established if the area is
two non-overlapping circles
of equal size.
The main free geometric parameter is the ratio of the pupil radii over the distance between
both circles.
Increasingly higher order aberrations---as defined for a virtual larger pupil
in which both pupils are embedded---are
fed into a Gram-Schmidt orthogonalization to
distill
one unique set of basis functions.
The key effort
is to work out the overlap integrals between a full set of
primitive basis functions, which are chosen to be products of powers of the distance from the mid-point
between both pupils by
azimuthal  functions of the Fourier type.
\end{abstract}

\maketitle
\section{Aim and Scope} 
Manufacturing schemes of lenses and  mirrors inevitably prefer circular
cross-sections of beams, and the associated description of functions
(aberrations)
defined across these fields calls for basis functions
on this circular support, the best-noted probably being the Zernike
functions
\cite{BhatiaPRSB65,NollJOSA66,ChongPR36,PrataAO28,SheppardAO43}.
Masking a central circular portion of a circular beam leads
to annular, ring-shaped
regions, for which orthogonal basis sets
are also established
\cite{DaiJOSAA24,HouAO45,MahajanJOSA71,WangAO19,SwantnerAO19}.

This work proceeds to the
task of defining such a basis set
for a two-beam interferometer,
in which the input pupil is defined
by two disconnected circular areas of
equal radius
\cite{HuJOSAA6}.

We define the area of integration in a global spherical coordinate system
centered in between the two apertures in Section \ref{sec.geom}.
Supposed anonymous functions defined over these apertures are
expanded with a separation ansatz as products of powers of the distance to the
origin of coordinates by the usual Fourier series in the azimuth, all
integrals over products of these can be reduced to a generic integral,
summarized in Section \ref{sec.A}.
The value of this article lies in the
the reduction formulas of two associated integrals
in two appendices.
Section \ref{sec.Gschm} proceeds with an application, the re-orthogonalization
of the Zernike basis functions---defined over the larger area that encompasses
both circles---with respect to the two circular regions that define
the binocular pupil.

\section{Binocular Geometry} 
\label{sec.geom}
\begin{figure}
\includegraphics[scale=0.7]{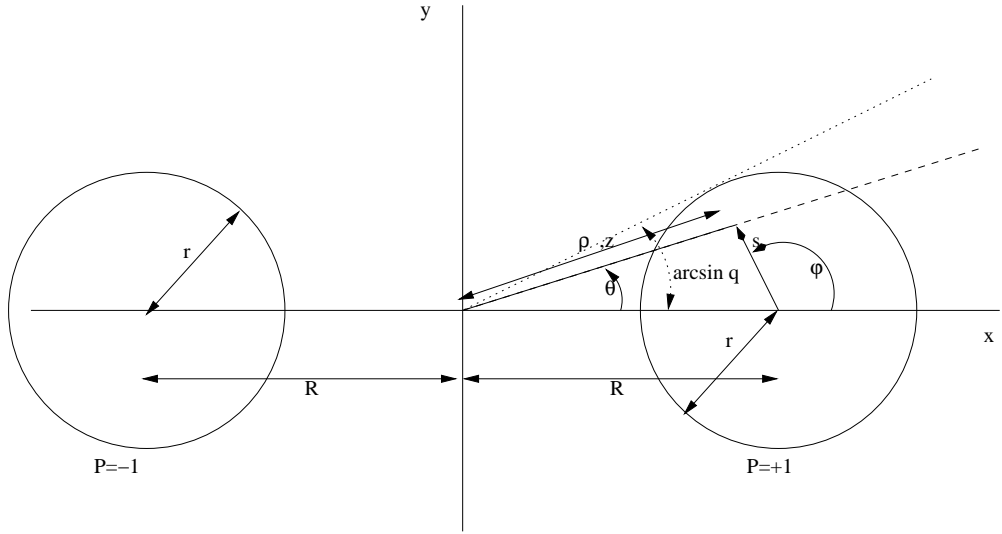}
\caption{The two-dimensional area and the Cartesian and
polar coordinates covered by the two circular sub-pupils.
The center of symmetry and origin of the global circular coordinates is at $x=y=0$.
In the first quadrant, the limit $\theta \le \arcsin q$ of integration is given by
the tangent of a radial vector on the pupil with index $P=+1$ along the dotted line.
For general $\theta$,
the dashed line illustrates how the radial vector intersects the front
and rear side of the circle, associated with the two signs in (\ref{eq.zofphi}).
}
\label{fig.geo}
\end{figure}

Orthogonality of functions $f_k$ and $f_l$ over two-dimensional areas is defined through
their product integrated over the area
\begin{equation}
(f_k,f_l)=\iint f_k^* f_l d\Omega \sim \delta_{kl}.
\label{eq.ortho}
\end{equation}
In Cartesian coordinates $x$ and $y$, or circular coordinates with distance
$\rho$ to the origin and azimuth $\theta$,
\begin{equation}
x=\rho\cos\theta ;\quad y=\rho\sin\theta,
\end{equation}
the differential
is $d\Omega=dx\,dy$ or $d\Omega=\rho\,d\rho\,d\theta$.
This manuscript deals with two-dimensional areas that are the sum
of the interior of two circular pupils represented by
\begin{equation}
(x\pm R)^2+y^2\le r^2,
\end{equation}
where $R$ is half the distance between the two pupil centers, where $2R$ is
the interferometric baseline, and $r$ is each pupil's radius (Fig.\ \ref{fig.geo}).
This transforms the integral operator into a sum over both circles,
\begin{equation}
\iint d\Omega =
\int_{R-r}^{R+r}\rho d\rho
\left[
\int_{2\rho R\cos\theta \le r^2-R^2-\rho^2} d\theta
+
\int_{2\rho R\cos\theta \ge \rho^2-r^2+R^2} d\theta
\right]
.
\end{equation}
Scaling distances in units of $R$, $z\equiv \rho/R$, leaves one
essential shape parameter, $q\equiv r/R$.
To avoid double-counting of areas,
these must not overlap:
\begin{equation}
\quad 0\le q\le 1.
\end{equation}
The range of radial distances that lie inside the circle centered at $x=+R$
for some fixed direction $\theta$ is
\begin{equation}
z
= \cos\theta \pm\sqrt{q^2-\sin^2\theta}
,
\label{eq.zofphi}
\end{equation}
and for the other one centered at $x=-R$,
\begin{equation}
z
= -\cos\theta \pm\sqrt{q^2-\sin^2\theta}
.
\end{equation}
In each of the two equations, the lower sign connects to the intersection
with the circle rim that is closer to the origin, the upper sign to the intersection
with the farther one.

\section{Generic Integral} 
\label{sec.A}

At the heart of this work is performing overlap integrals
(\ref{eq.ortho}) in analytical terms over the area described above;
since this will be based on spanning the functional space with ``primitive'' basis
functions of the type
\begin{equation}
p_k \equiv p_{n_k,m_k}(z,\theta)\equiv z^{n_k}e^{im_k\theta},
\label{eq.pBas}
\end{equation}
the generic integral reads
\begin{equation}
A_{n,m}(q)\equiv
\int_{|\sin\theta|\le q}
d\theta
\int_{
\substack{
2z\cos\theta \le q^2-z^2-1 \\
2z \cos\theta \ge z^2-q^2+1
}
}
zdz
\,
z^ne^{im\theta}
,
\quad n,m=0,1,2,\ldots
\label{eq.Adef}
\end{equation}
$i$ is the imaginary unit, $m$ the azimuthal frequency, and $n$ the power
to build a complete radial basis.
With a sign tag $P$ defined for both circles,
\begin{equation}
P\equiv\left\{ \begin{array}{ll} +1, & |\theta|<\pi/2 , \\ -1, &
\pi/2 \le |\theta|\le \pi ,
\end{array}\right.
\label{eq.Tpar}
\end{equation}
the integration over $z^{n+1}$ may be executed,
\begin{multline}
(n+2)A_{n,m}(q)=
\int_{|\sin \theta|\le q}d\theta
\Big[\Big(P\cos\theta+\sqrt{q^2-\sin^2\theta}\Big)^{n+2}
-\Big(P\cos\theta-\sqrt{q^2-\sin^2\theta}\Big)^{n+2}\Big]e^{im\theta}
.
\end{multline}
The contribution from the imaginary part proportional to 
$\sin(m\theta)$ vanishes;
$e^{im\theta}$ can be replaced by $\cos(m\theta)$ in this equation.
Considering the coordinate transformation $y\leftrightarrow -y$
or $\theta \leftrightarrow -\theta$, the integral over all four quadrants
can be reduced to an integral over the first and second quadrant and a
factor of 2:
\begin{multline}
(n+2)A_{n,m}(q)=2\int_{\substack{|\sin \theta|\le q\\ 0\le \theta\le \pi}}
d\theta
\Big[\Big(P\cos\theta+\sqrt{q^2-\sin^2\theta}\Big)^{n+2}
-\Big(P\cos\theta-\sqrt{q^2-\sin^2\theta}\Big)^{n+2}\Big]
\cos(m\theta)
.
\end{multline}
Considering also the variable transformation $\theta \leftrightarrow \pi-\theta$,
the parities of the Chebyshev term $T_m(\cos\theta)=\cos(m\theta)$, of $\cos\theta$
and of $P$, this vanishes for odd $m$ and reduces to an integral over the first quadrant
for even $m$,
\begin{multline}
\frac{(n+2)A_{n,m}(q)}{4}
=\int_{\substack{\sin \theta\le q\\ 0\le \theta\le \pi/2}}
d\theta
\Big[\Big(\cos\theta+\sqrt{q^2-\sin^2\theta}\Big)^{n+2}
-\Big(\cos\theta-\sqrt{q^2-\sin^2\theta}\Big)^{n+2}\Big]
\cos(m\theta)
.
\end{multline}
If we define the integrals
\begin{equation}
B_{n+2,m}^\pm(q)
\equiv
\int_0^{\arcsin q}d\theta
\left(\cos\theta\pm\sqrt{q^2-\sin^2\theta}\right)^{n+2}T_m(\cos\theta)
,
\label{eq.Bdef}
\end{equation}
this can be rephrased
\begin{equation}
\frac{(n+2)A_{n,m}(q)}{4}
=
\left\{
\begin{array}{ll}
0 & , m=1,3,5,7,\ldots \\
B_{n+2,m}^+(q)-B_{n+2,m}^-(q) & , m=0,2,4,6,\ldots
\end{array}
\right .
.
\label{eq.AofB}
\end{equation}

Table \ref{tab.A} summarizes the results obtained in
Appendix \ref{sec.B}
as series expansions of 
$A_{n,m}(q)$ for small $n$ and $m$.

\begin{longtable}[l]{p{0.5cm}p{0.5cm}p{16cm}}
$n$ & $m$ & $A_{n,m}(q)/\pi$ \\
\hline
0&0&$ +2q^2$\\

0&2&$ +2q^2 -q^4$\\

0&4&$ +2q^2 -4q^4 +2q^6$\\

0&6&$ +2q^2 -9q^4 +12q^6 -5q^8$\\

0&8&$ +2q^2 -16q^4 +40q^6 -40q^8 +14q^{10}$\\

0&10&$ +2q^2 -25q^4 +100q^6 -175q^8 +140q^{10} -42q^{12}$\\

1&0&$ +2q^2 +1/4q^4 +1/96q^6 +1/512q^8 +5/8192q^{10} +49/196608q^{12}$\\
&&$ +63/524288q^{14} +1089/16777216q^{16}+\ldots$\\

1&2&$ +2q^2 -3/4q^4 +3/32q^6 +5/512q^8 +21/8192q^{10} +63/65536q^{12}$\\
&&$ +231/524288q^{14} +3861/16777216q^{16}+\ldots$\\

1&4&$ +2q^2 -15/4q^4 +75/32q^6 -175/512q^8 -315/8192q^{10} -693/65536q^{12}$\\
&&$ -2145/524288q^{14} -32175/16777216q^{16}+\ldots$\\

1&6&$ +2q^2 -35/4q^4 +1225/96q^6 -3675/512q^8 +8085/8192q^{10}$\\
&&$ +7007/65536q^{12} +15015/524288q^{14} +182325/16777216q^{16}+\ldots$\\

2&0&$ +2q^2 +q^4$\\

2&2&$ +2q^2$\\

2&4&$ +2q^2 -3q^4 +2q^6 -1/2q^8$\\

2&6&$ +2q^2 -8q^4 +12q^6 -8q^8 +2q^{10}$\\

2&8&$ +2q^2 -15q^4 +40q^6 -50q^8 +30q^{10} -7q^{12}$\\

2&10&$ +2q^2 -24q^4 +100q^6 -200q^8 +210q^{10} -112q^{12} +24q^{14}$\\

3&0&$ +2q^2 +9/4q^4 +3/32q^6 +1/512q^8 +9/40960q^{10} +3/65536q^{12}$\\
&&$ +7/524288q^{14} +81/16777216q^{16}+\ldots$\\

3&2&$ +2q^2 +5/4q^4 -5/32q^6 +5/512q^8 +5/8192q^{10} +7/65536q^{12}$\\
&&$ +15/524288q^{14} +165/16777216q^{16}+\ldots$\\

3&4&$ +2q^2 -7/4q^4 +35/32q^6 -175/512q^8 +245/8192q^{10} +147/65536q^{12}$\\
&&$ +231/524288q^{14} +2145/16777216q^{16}+\ldots$\\

3&6&$ +2q^2 -27/4q^4 +315/32q^6 -3675/512q^8 +19845/8192q^{10}$\\
&&$ -14553/65536q^{12} -9009/524288q^{14} -57915/16777216q^{16}+\ldots$\\

4&0&$ +2q^2 +4q^4 +2/3q^6$\\

4&2&$ +2q^2 +3q^4$\\

4&4&$ +2q^2$\\

4&6&$ +2q^2 -5q^4 +20/3q^6 -5q^8 +2q^{10} -1/3q^{12}$\\

4&8&$ +2q^2 -12q^4 +30q^6 -40q^8 +30q^{10} -12q^{12} +2q^{14}$\\

4&10&$ +2q^2 -21q^4 +84q^6 -175q^8 +210q^{10} -147q^{12} +56q^{14} -9q^{16}$\\

5&0&$ +2q^2 +25/4q^4 +75/32q^6 +25/512q^8 +5/8192q^{10} +3/65536q^{12}$\\
&&$ +25/3670016q^{14} +25/16777216q^{16}+\ldots$\\

5&2&$ +2q^2 +21/4q^4 +35/32q^6 -35/512q^8 +21/8192q^{10} +7/65536q^{12}$\\
&&$ +7/524288q^{14} +45/16777216q^{16}+\ldots$\\

5&4&$ +2q^2 +9/4q^4 -21/32q^6 +105/512q^8 -315/8192q^{10} +147/65536q^{12}$\\
&&$ +63/524288q^{14} +297/16777216q^{16}+\ldots$\\

5&6&$ +2q^2 -11/4q^4 +99/32q^6 -1155/512q^8 +8085/8192q^{10}$\\
&&$ -14553/65536q^{12} +7623/524288q^{14} +14157/16777216q^{16}+\ldots$\\

6&0&$ +2q^2 +9q^4 +6q^6 +1/2q^8$\\

6&2&$ +2q^2 +8q^4 +4q^6$\\

6&4&$ +2q^2 +5q^4$\\

6&6&$ +2q^2$\\

6&8&$ +2q^2 -7q^4 +14q^6 -35/2q^8 +14q^{10} -7q^{12} +2q^{14} -1/4q^{16}$\\

6&10&$ +2q^2 -16q^4 +56q^6 -112q^8 +140q^{10} -112q^{12} +56q^{14} -16q^{16}$\\

7&0&$ +2q^2 +49/4q^4 +1225/96q^6 +1225/512q^8 +245/8192q^{10}$\\
&&$ +49/196608q^{12} +7/524288q^{14} +25/16777216q^{16}+\ldots$\\

7&2&$ +2q^2 +45/4q^4 +315/32q^6 +525/512q^8 -315/8192q^{10} +63/65536q^{12}$\\
&&$ +15/524288q^{14} +45/16777216q^{16}+\ldots$\\

7&4&$ +2q^2 +33/4q^4 +99/32q^6 -231/512q^8 +693/8192q^{10} -693/65536q^{12}$\\
&&$ +231/524288q^{14} +297/16777216q^{16}+\ldots$\\

7&6&$ +2q^2 +13/4q^4 -143/96q^6 +429/512q^8 -3003/8192q^{10}$\\
&&$ +7007/65536q^{12} -9009/524288q^{14} +14157/16777216q^{16}+\ldots$\\

8&0&$ +2q^2 +16q^4 +24q^6 +8q^8 +2/5q^{10}$\\

8&2&$ +2q^2 +15q^4 +20q^6 +5q^8$\\

8&4&$ +2q^2 +12q^4 +10q^6$\\

8&6&$ +2q^2 +7q^4$\\

8&8&$ +2q^2$\\

8&10&$ +2q^2 -9q^4 +24q^6 -42q^8 +252/5q^{10} -42q^{12} +24q^{14} -9q^{16}$\\

9&0&$ +2q^2 +81/4q^4 +1323/32q^6 +11025/512q^8 +19845/8192q^{10}$\\
&&$ +1323/65536q^{12} +63/524288q^{14} +81/16777216q^{16}+\ldots$\\

9&2&$ +2q^2 +77/4q^4 +1155/32q^6 +8085/512q^8 +8085/8192q^{10}$\\
&&$ -1617/65536q^{12} +231/524288q^{14} +165/16777216q^{16}+\ldots$\\

9&4&$ +2q^2 +65/4q^4 +715/32q^6 +2145/512q^8 -3003/8192q^{10}$\\
&&$ +3003/65536q^{12} -2145/524288q^{14} +2145/16777216q^{16}+\ldots$\\

9&6&$ +2q^2 +45/4q^4 +195/32q^6 -715/512q^8 +3861/8192q^{10}$\\
&&$ -9009/65536q^{12} +15015/524288q^{14} -57915/16777216q^{16}+\ldots$\\

10&0&$ +2q^2 +25q^4 +200/3q^6 +50q^8 +10q^{10} +1/3q^{12}$\\

10&2&$ +2q^2 +24q^4 +60q^6 +40q^8 +6q^{10}$\\

10&4&$ +2q^2 +21q^4 +42q^6 +35/2q^8$\\

10&6&$ +2q^2 +16q^4 +56/3q^6$\\

10&8&$ +2q^2 +9q^4$\\

10&10&$ +2q^2$\\
\caption{Series expansion of $A_{n,m}(q)/\pi$. It is a polynomial in $q^2$
if $n+m$ is even, else shown in truncated form up to $O(q^{16})$.
}
\label{tab.A}
\end{longtable}

To lowest order in $q$,
$A_{n,m}$ equals $2\pi q^2$,
the total
area of both circles. This is expected, because for circles far
away from their common origin at
$z=0$,
the relative strength
of the variation introduced
by the power $n$ and by the modulation $\propto \cos(m\theta)$
of the function in the integral kernel
loses importance.

\section{Gram-Schmidt Orthogonalization} 
\label{sec.Gschm}
\subsection{Procedure.}

The integral evaluated in Section \ref{sec.A} allows to calculate the overlap
integral (inner product) between any two functions expressed as linear combinations
(``contractions'') of ``primitive'' basis functions of the form (\ref{eq.pBas})
in the global circular coordinate system centered at the middle between
the two sub-pupils, because the overlap between two of these is 
\begin{equation}
(p_k,p_l)
\equiv \iint p_k^*p_l d\Omega
=\iint
z^{n_k}e^{-im_k\theta}z^{n_l}e^{im_l\theta}z dz d\theta
=
A_{n_k+n_l,|m_k-m_l|}(q).
\label{eq.Odef}
\end{equation}
The quickest, obvious way of obtaining some orthogonal basis set from any set of contracted
primitive basis function is to diagonalize the
overlap matrix
containing all the overlap integrals between pairs of these basis functions \cite{DaiOL32}.
To end up with some standardization of these orthogonal bases, we use the Gram-Schmidt
procedure, which builds this set $\{f_k\}$ incrementally. At each step of
the procedure, an ansatz
\begin{equation}
f_{k+1}(z,\theta)=\beta_{k+1}\left[g_{k+1}(z,\theta)
+\sum_{l=1}^k \gamma_{k+1,l}f_l(z,\theta)\right]
\end{equation}
is made for the next, $(k+1)$st additional basis function $f_{k+1}$, given
a seed function $g_{k+1}(z,\theta)$ plus the $f_l$ generated by the earlier steps.
Essentially, the projections of the seed along all earlier directions
are subtracted, and the residual is normalized to unity.
The request of orthogonality
\begin{equation}
\iint f_k^*f_l d\Omega = (f_k,f_l)=\delta_{kl},\quad k,l=1,\ldots k+1,
\end{equation}
means the projection coefficients $\gamma$ can be calculated from the $k$
overlaps between the seed and the earlier basis functions,
\begin{equation}
\gamma_{k+1,l}= -(g_{k+1},f_l),\quad l=1,\ldots k.
\end{equation}
The normalization
$\beta_{k+1}$ is finally computed from
the self-overlap of the seed and the sum over the squared $\gamma$,
\begin{equation}
1=\beta_{k+1}^2\left[(g_{k+1},g_{k+1})-\sum_{l=1}^k \gamma_{k+1,l}^2\right].
\end{equation}
The Gram-Schmidt methodology is well known
in the literature for different geometric shapes of pupils
\cite{SwantnerAO33,UptonOL29,MahajanJOSAA24}.

Still, the procedure establishes different basis sets depending on the
order in which the seeds $g$ are fed into the procedure, and depending on which
functional subspaces they span.

\subsection{Zernike Seeds.}
\label{sec.f}
To define a unique set $\{f_j\}$
of functions orthogonal over the binocular pupil, we may choose the real and
imaginary parts of the primitives (\ref{eq.pBas}) in increasing order of
complexity, ie, increasing order of aberration and increasing $n$ and $m$, as the seeds,
as if one would subduce the Zernike polynomials $Z_j(z,\theta)$ over the full super-pupil
of radius
$R$ (normalized to $z=1$)
in the Noll order of indexing \cite{NollJOSA66} into
the two sub-apertures.
(We identify the variable $z$ with the radial variable of the Zernike polynomials,
although the corresponding Zernike radius would need to be $1+q$, not 1, to
cover both sub-apertures in full.)
There is an ``outer'' loop over $n=0,1,2,\ldots$
and an ``inner'' loop over $m=n\pmod 2,\ldots n$, considering only even $n-m$:
\begin{eqnarray}
g_1=1;\quad g_2=z\cos\theta;\quad g_3=z\sin\theta;
\label{eq.g1}
\\
g_4=2z^2-1;\quad g_5=z^2\sin(2\theta);\quad g_6=z^2\cos(2\theta);
\\
g_7=(3z^3-2z)\sin\theta;\quad g_8=(3z^3-2z)\cos\theta;\quad
g_9=z^3\sin(3\theta);\quad g_{10}=z^3\cos(3\theta);\quad \ldots.
\label{eq.g9}
\end{eqnarray}
For this choice, the $n+1$ basis functions from
$g_{1+n(n+1)/2}$ up to and including $g_{(n+1)(n+2)/2}$ are associated with
a  polynomial of order $n$ in $z$.
The arithmetic remains real-valued, because the ``atoms'' of the seeds
are the separated real and imaginary part of (\ref{eq.pBas}).
(\ref{eq.Odef}) is
split into
\begin{eqnarray}
\iint z^{n_k}\cos(m_k\theta)z^{n_l}\cos(m_l\theta)d\Omega
&=&
\frac{1}{2}A_{n_k+n_l,m_k+m_l}
+\frac{1}{2}A_{n_k+n_l,|m_k-m_l|} ;
\\
\iint z^{n_k}\sin(m_k\theta)z^{n_l}\sin(m_l\theta)d\Omega
&=&
\frac{1}{2}A_{n_k+n_l,|m_k-m_l|}
-\frac{1}{2}A_{n_k+n_l,m_k+m_l} ;
\\
\iint z^{n_k}\cos(m_k\theta)z^{n_l}\sin(m_l\theta)d\Omega
&=&
0.
\end{eqnarray}

The first basis functions created with this recipe are discussed shortly
in analytical form.
We start with the ``global common piston'' $g_1$ which just needs to be normalized:
\begin{equation}
f_1=
\frac{1}{q}\frac{1}{\sqrt{2\pi}}
.
\label{eq.f1}
\end{equation}
Next we feed what represents most of the differential piston, $g_2$, which turns
out to be already orthogonal to $f_1$
and only needs to be normalized,
\begin{equation}
f_2=
\frac{1}{q}\sqrt{\frac{2}{\pi(4+q^2)}} z \cos\theta
.
\end{equation}
Next we feed $g_3$, some
common sideways tilt perpendicular to
the baseline between the two pupils,
\begin{equation}
f_3=
\frac{1}{q^2}
\sqrt{\frac{2}{\pi}}
 z\sin\theta
.
\end{equation}
The first case of nonzero overlap with an earlier basis function occurs when
we use $g_4$ as a seed (some
nodding tilt between the sub-pupils
along the baseline), which has a nonzero component along $f_1$:
\begin{equation}
f_4=
\frac{1}{q^2}\sqrt{\frac{6}{\pi(12+q^2)}}\left(z^2
-\frac{2+q^2}{2}\right)
.
\end{equation}
Feeding $g_5$ to $g_8$ we get
\begin{equation}
f_5=
\frac{1}{q^2}
\sqrt{\frac{3}{\pi(6+q^2)}}
 z^2 \sin(2\theta)
.
\end{equation}
\begin{equation}
f_6=
\frac{1}{q}
\sqrt{\frac{3}{\pi(18+q^2)(12+q^2)}}
\left[
\frac{12+q^2}{q^2} z^2 \cos(2\theta)
-
\frac{12}{q^2} z^2
+
5
\right]
.
\end{equation}
\begin{equation}
f_7=
\frac{2}{q^3}
\sqrt{\frac{1}{\pi(12+q^2)}}
\left[
3 z^3 \sin\theta
-
(3+2q^2)z \sin\theta
\right]
.
\end{equation}
\begin{eqnarray}
f_8&=&
\frac{2}{q^2}
\sqrt{\frac{1}{\pi Q_8(4+q^2)}}
\left[
3(4+q^2) z^3 \cos\theta
-
(12+21q^2+2q^4)
z \cos\theta
\right]
.
\end{eqnarray}
The abbreviation $Q_8\equiv 288+48q^2+48q^4+q^6$ is used. Continuing with $g_9$,
\begin{equation}
f_9=
\frac{2}{q^2}
\sqrt{\frac{1}{\pi(24+q^2)(12+q^2)}}
\left[
\frac{12+q^2}{q^2} z^3 \sin(3\theta)
+
21 z \sin(\theta)
-
\frac{36}{q^2} z^3 \sin\theta
\right]
.
\end{equation}
\begin{multline}
f_{10}=
\frac{2}{q}
\sqrt{\frac{1}{\pi Q_8(2304+704q^2+480q^4+60q^6+q^8)}}
\\
\times
\Big[
\frac{Q_8}{q^2} z^3 \cos(3\theta)
+
21 (4+2q+q^2)(4-2q+q^2) z \cos(\theta)
-
\frac{12(24+8q^2+3q^4)}{q^2} z^3 \cos\theta
\Big]
.
\end{multline}
\begin{multline}
f_{11}=
\frac{1}{q^3}
\sqrt{\frac{10}{\pi Q_{11}(18+q^2)}}
\\
\times
\Big[
3(18+q^2) z^4
-
3 (24+26q^2+q^4) z^2
+
\frac{108+30q^2+36q^4+q^6}{2}
-
12(3+q^2) z^2 \cos(2\theta)
\Big]
.
\end{multline}
The abbreviation $Q_{11}\equiv 2880+240q^2+78q^4+q^6$ is used.
\begin{multline}
f_{12}=
\frac{1}{q^2}
\sqrt{\frac{5}{\pi Q_{11}(64000+10240q^2+1840q^4+128q^6+q^8)}}
\\
\times
\Big[
\frac{4Q_{11}}{q^2} z^4\cos(2\theta)
-
\frac{3(3840+3200q^2+400q^4+84q^6+q^8)}{q^2} z^2\cos(2\theta)
-
21 (160+108q^2+4q^4+q^6)
\\
+
\frac{12 (960+1040q^2+38q^4+13q^6)}{q^2} z^2
-
\frac{60(192+8q^2+3q^4)}{q^2} z^4
\Big]
.
\end{multline}
\begin{multline}
f_{13}=
\frac{1}{q^3}
\sqrt{\frac{5}{\pi (640+80q^2+56q^4+q^6)(6+q^2)}}
\Big[
4(6+q^2) z^4\sin(2\theta)
-
3(8+12q^2+q^4) z^2\sin(2\theta)
\Big]
.
\label{eq.f13}
\end{multline}

\begin{figure}
\includegraphics[scale=0.42]{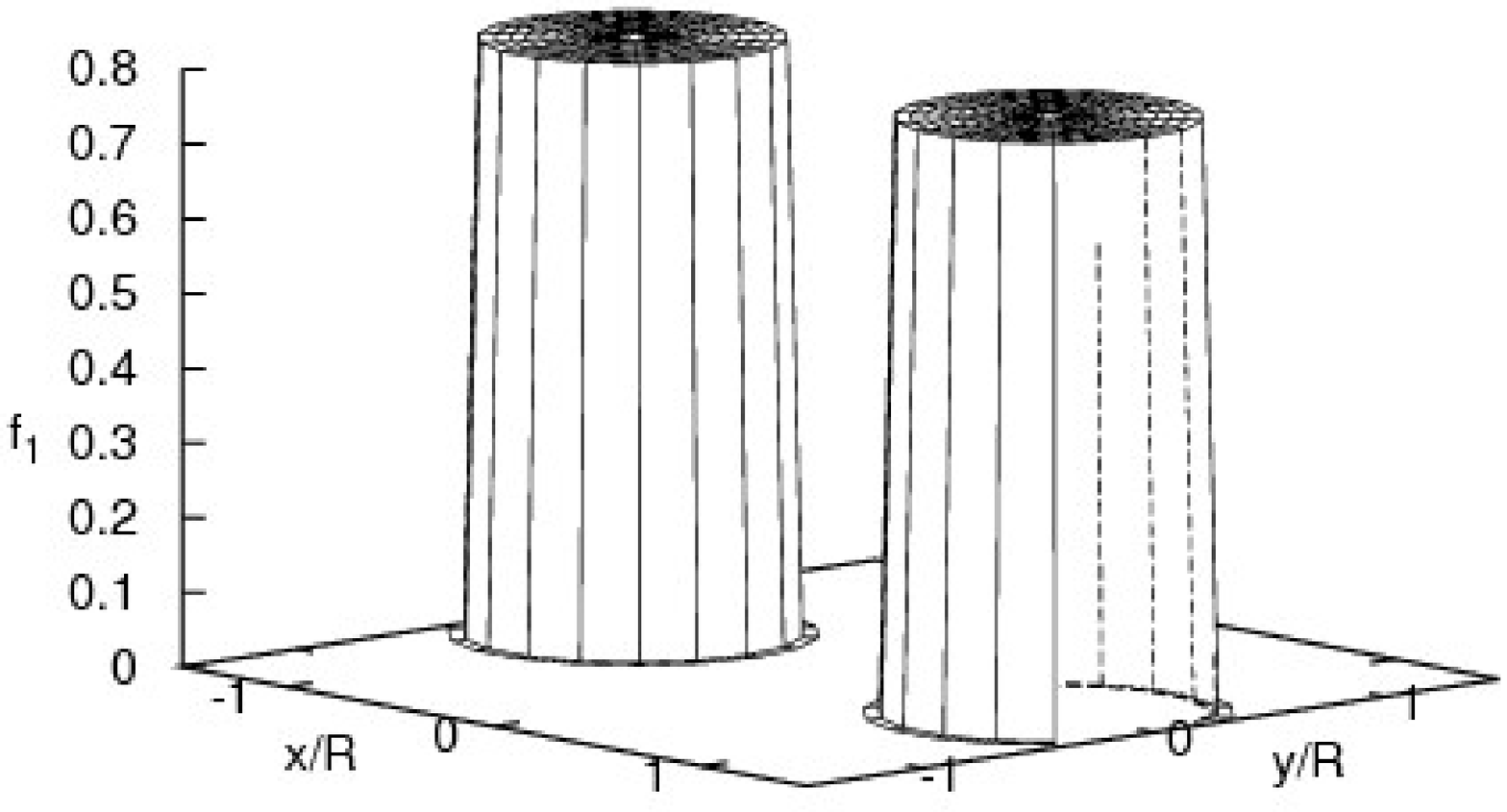}
\includegraphics[scale=0.42]{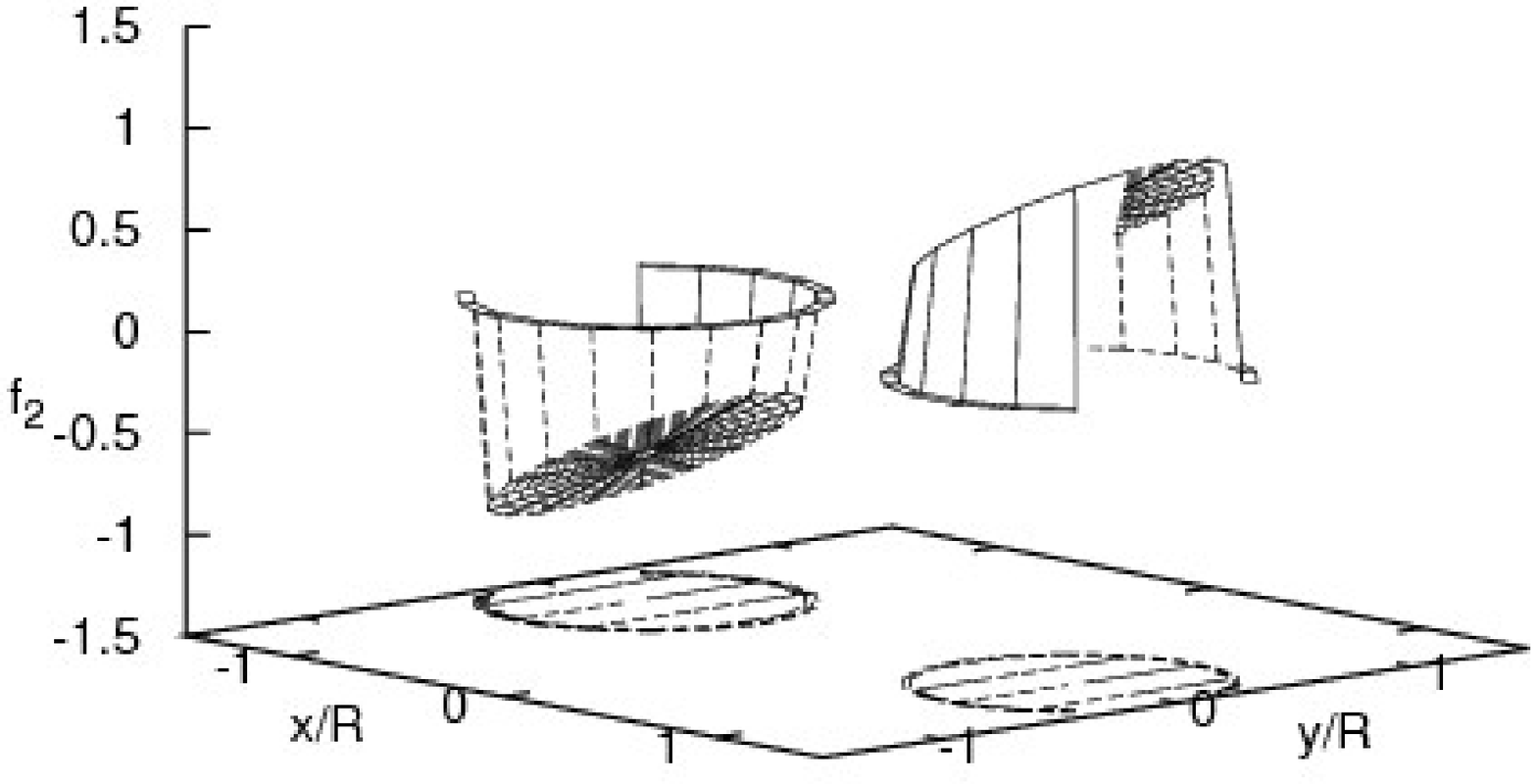}
\includegraphics[scale=0.42]{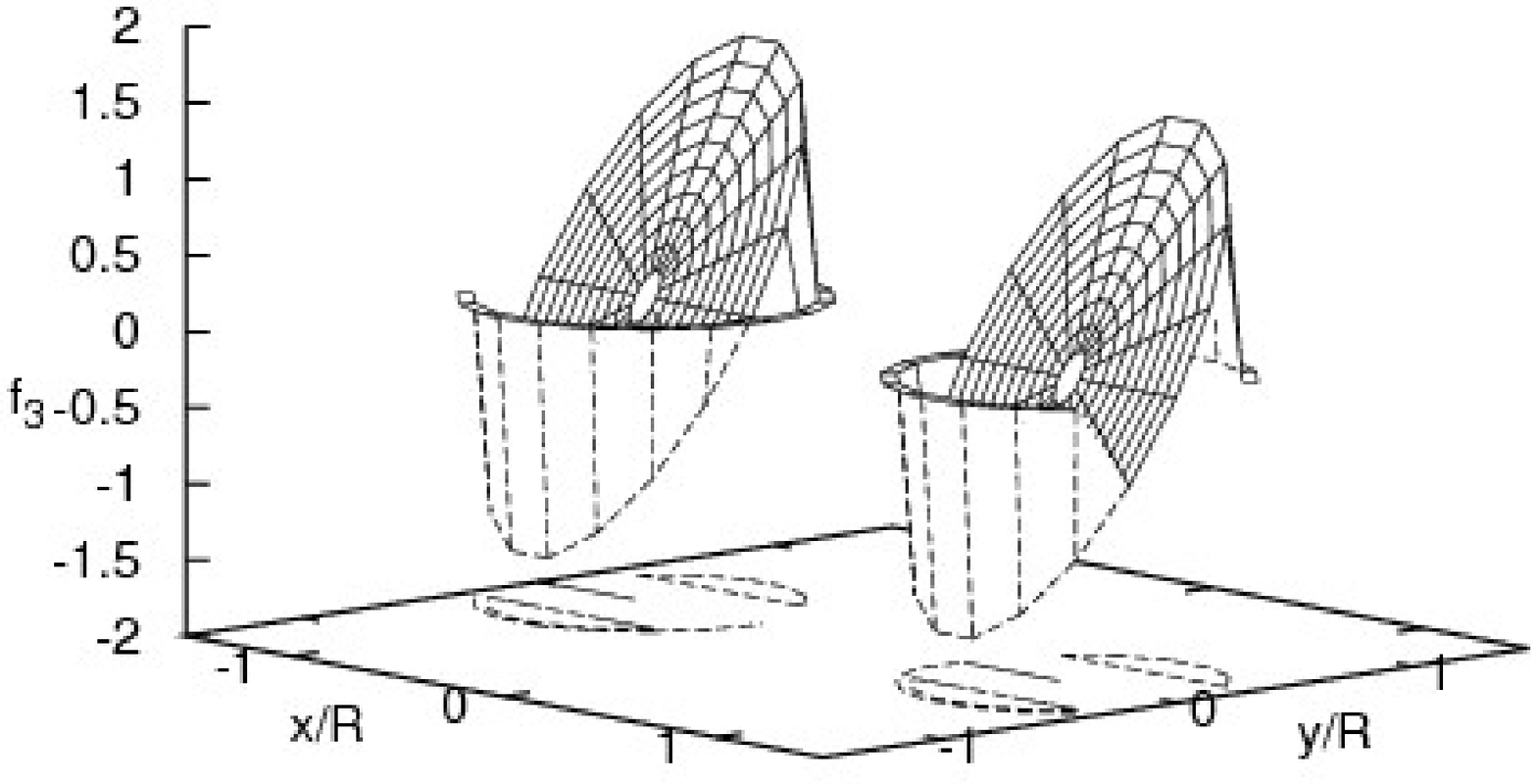}
\includegraphics[scale=0.42]{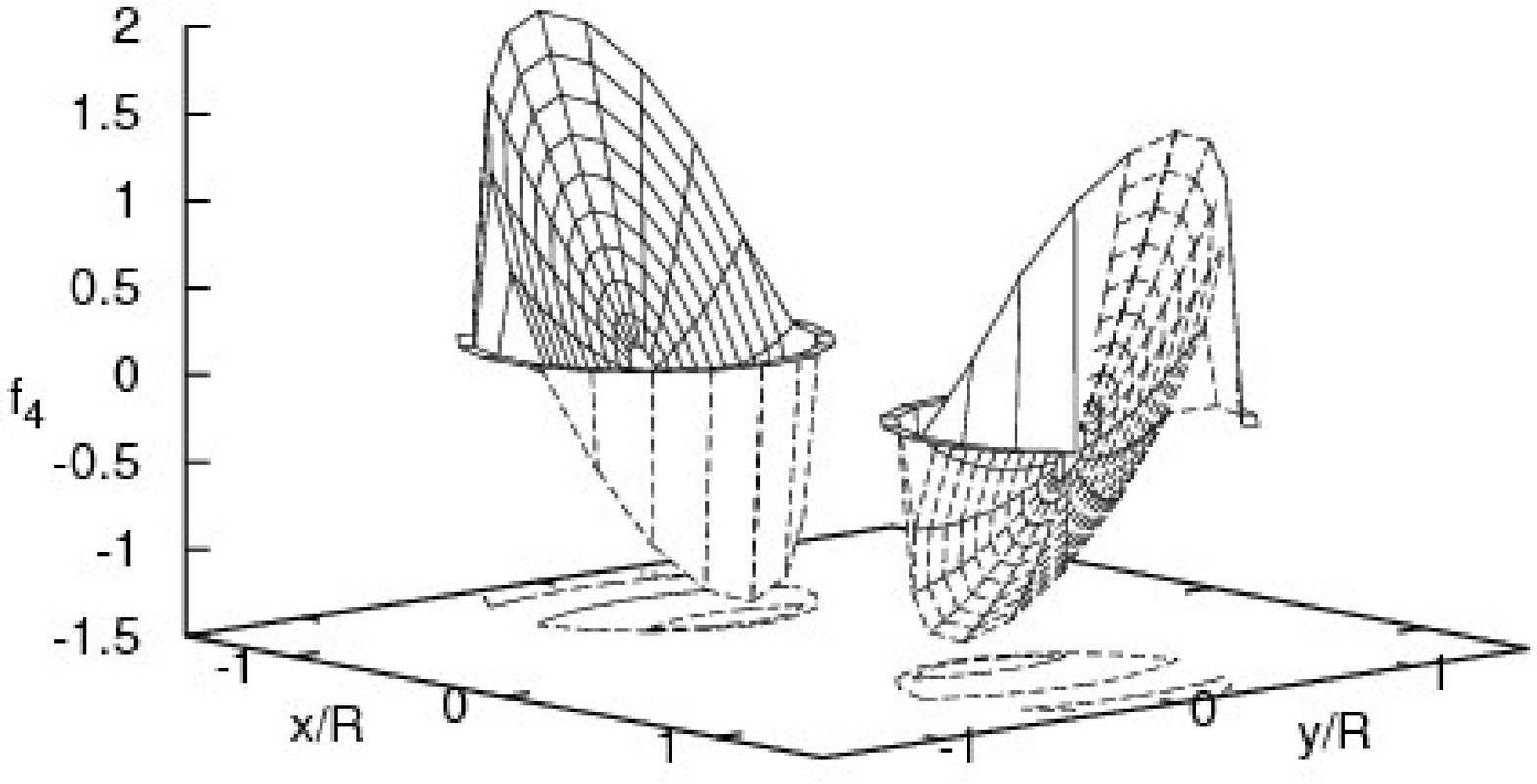}
\includegraphics[scale=0.42]{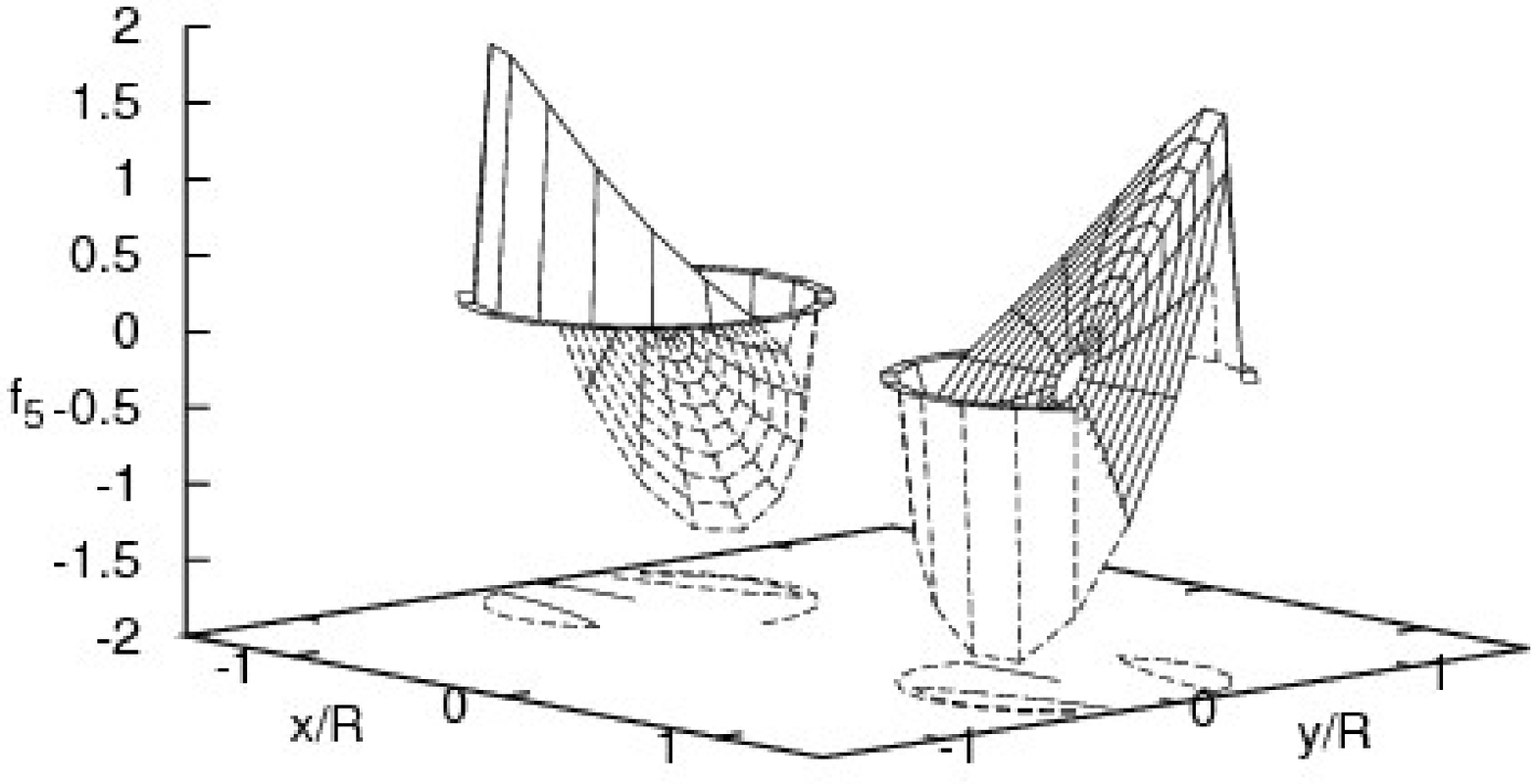}
\includegraphics[scale=0.42]{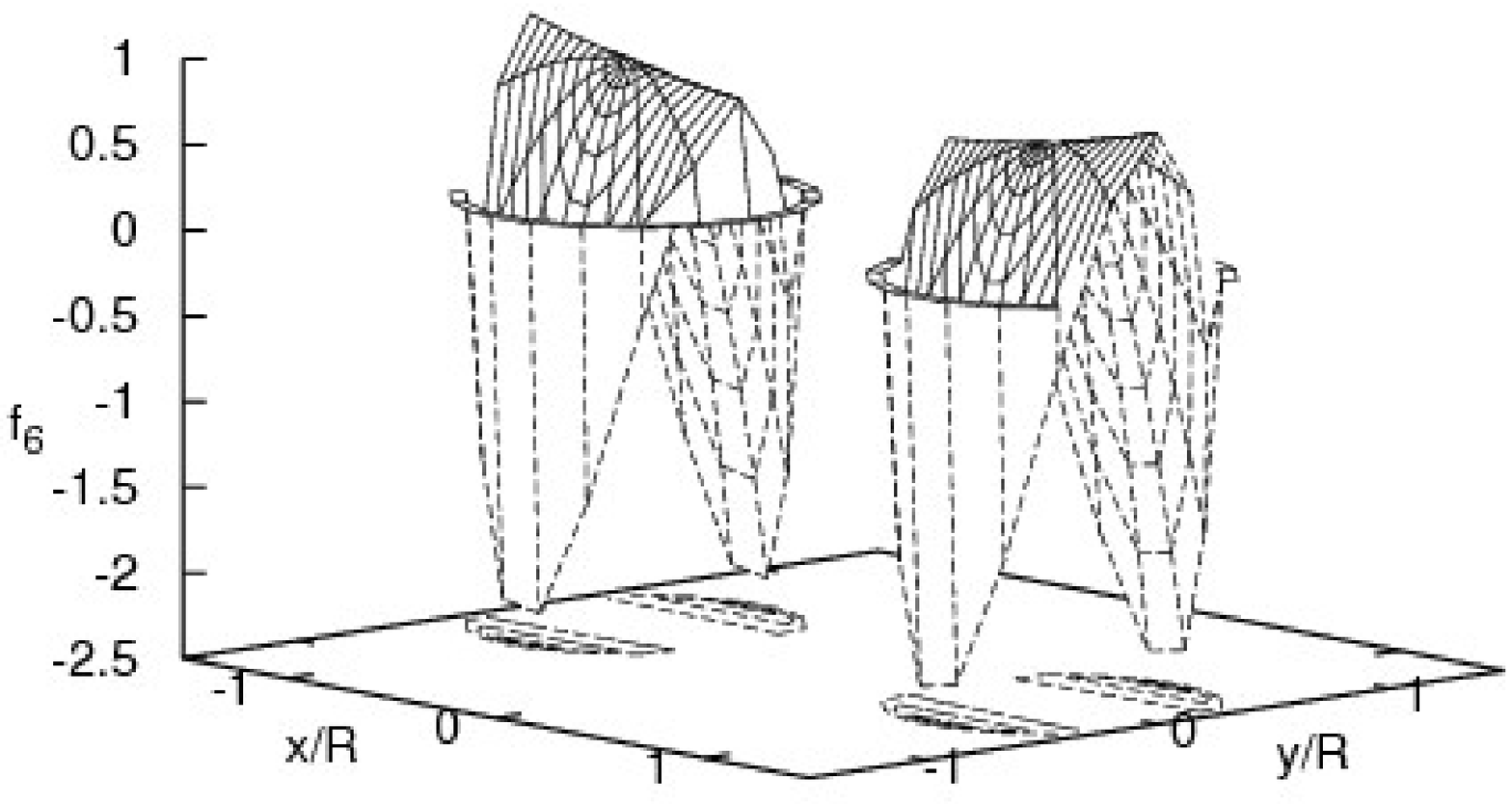}
\caption{The basis functions $f_1$ to $f_6$ for $q=1/2$.
The level of $f_l=0$ is indicated with small horse-shoe rims
around both sub-pupils.
}
\label{fig.displ1}
\end{figure}

\begin{figure}
\includegraphics[scale=0.42]{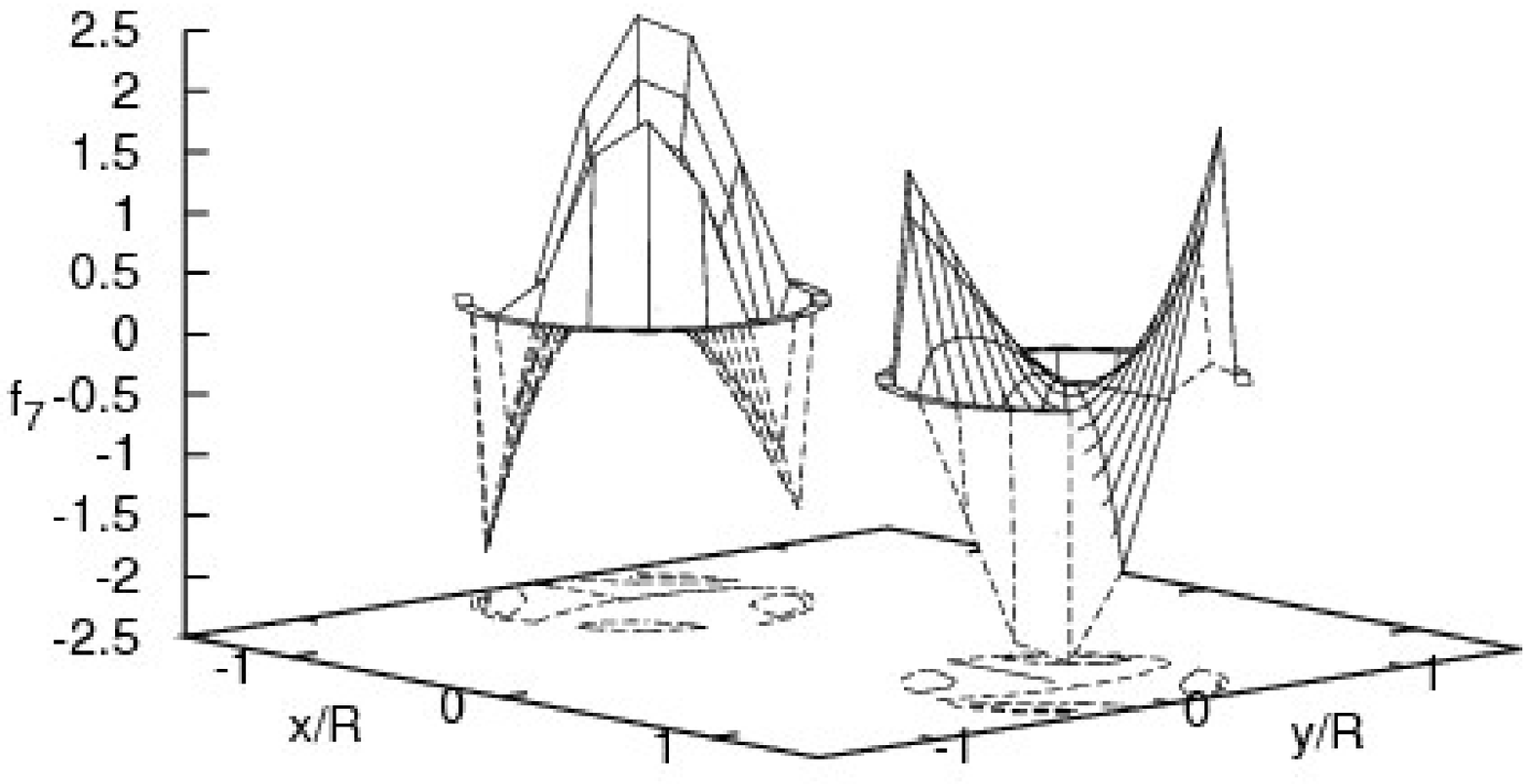}
\includegraphics[scale=0.42]{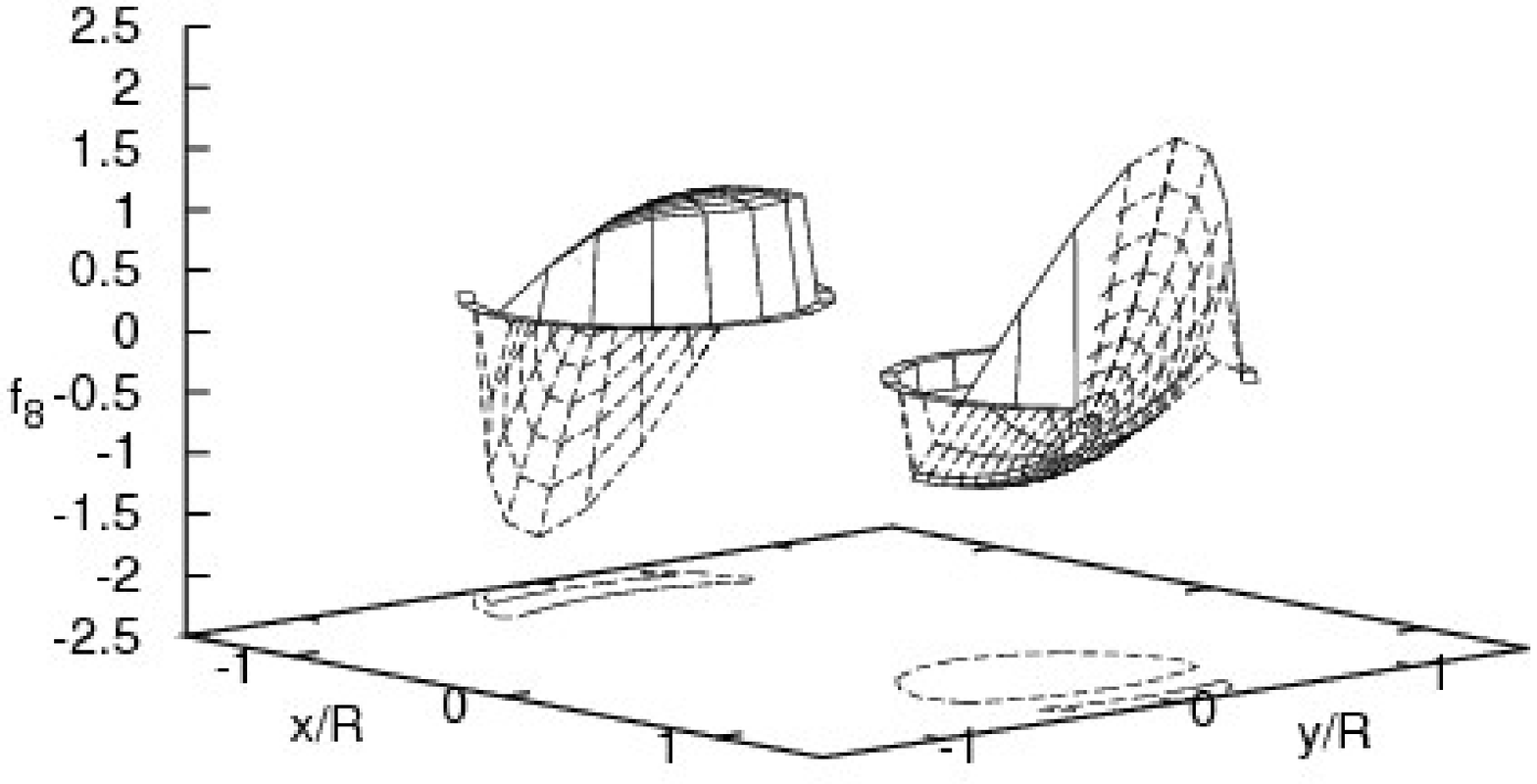}
\includegraphics[scale=0.42]{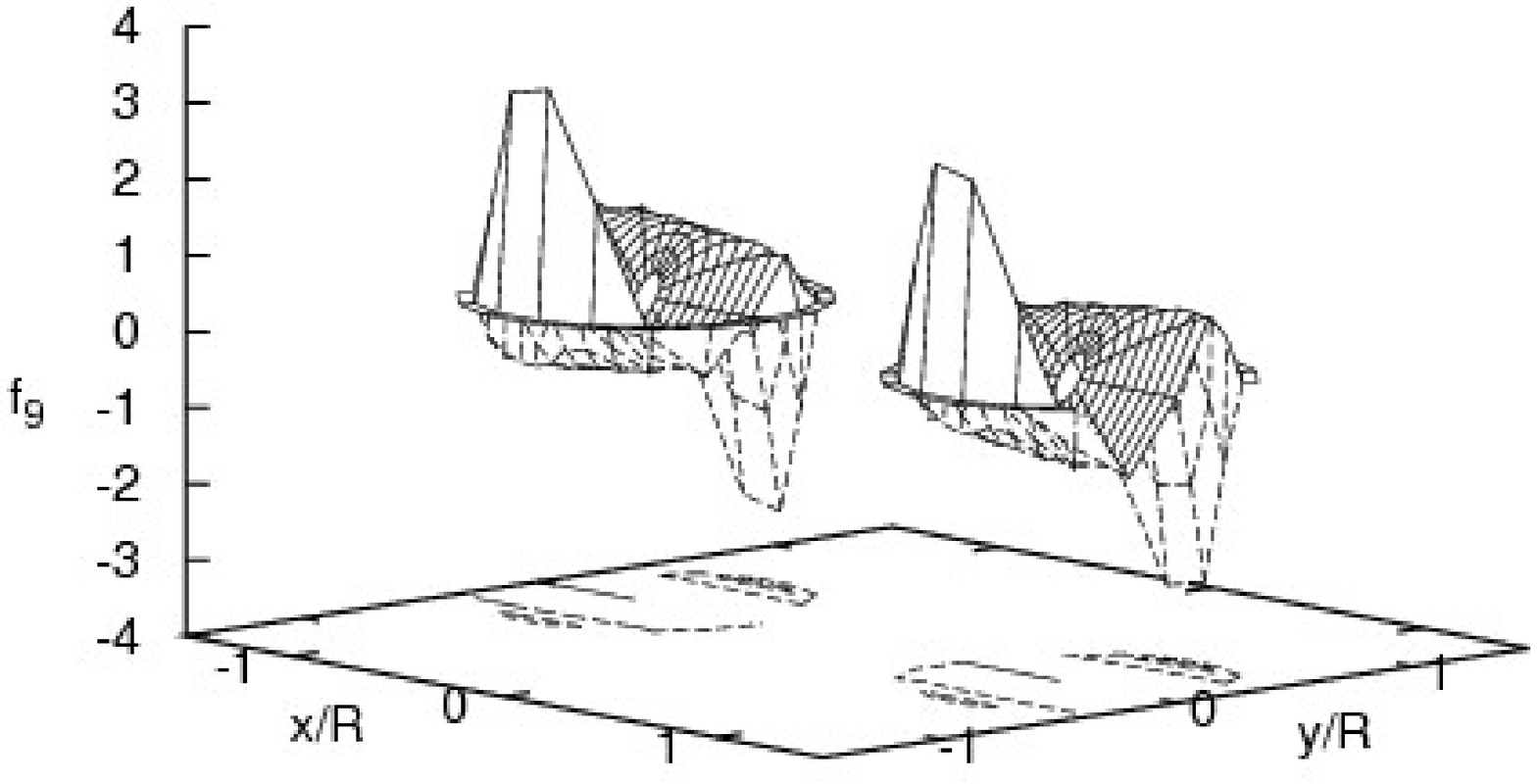}
\includegraphics[scale=0.42]{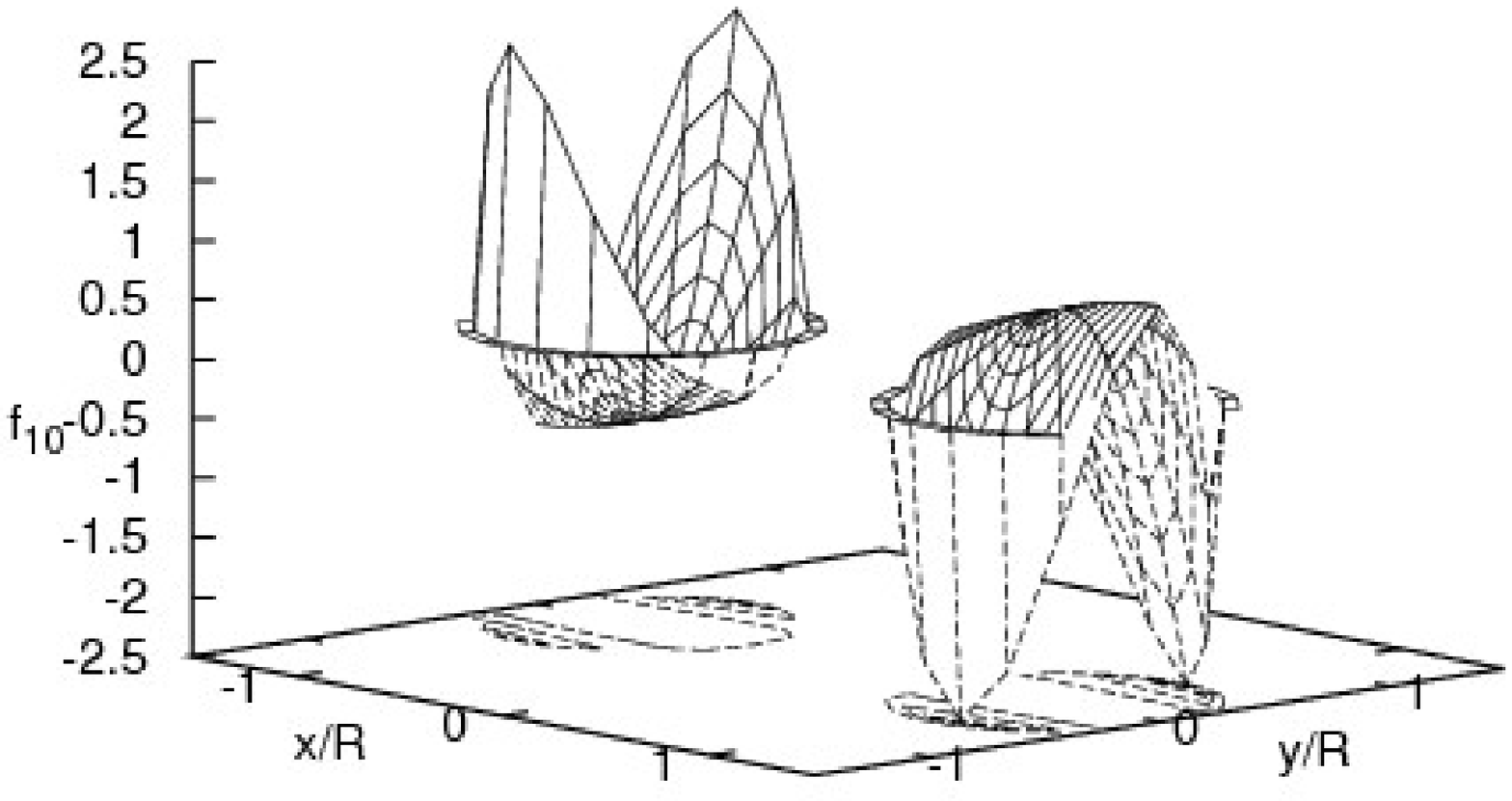}
\includegraphics[scale=0.42]{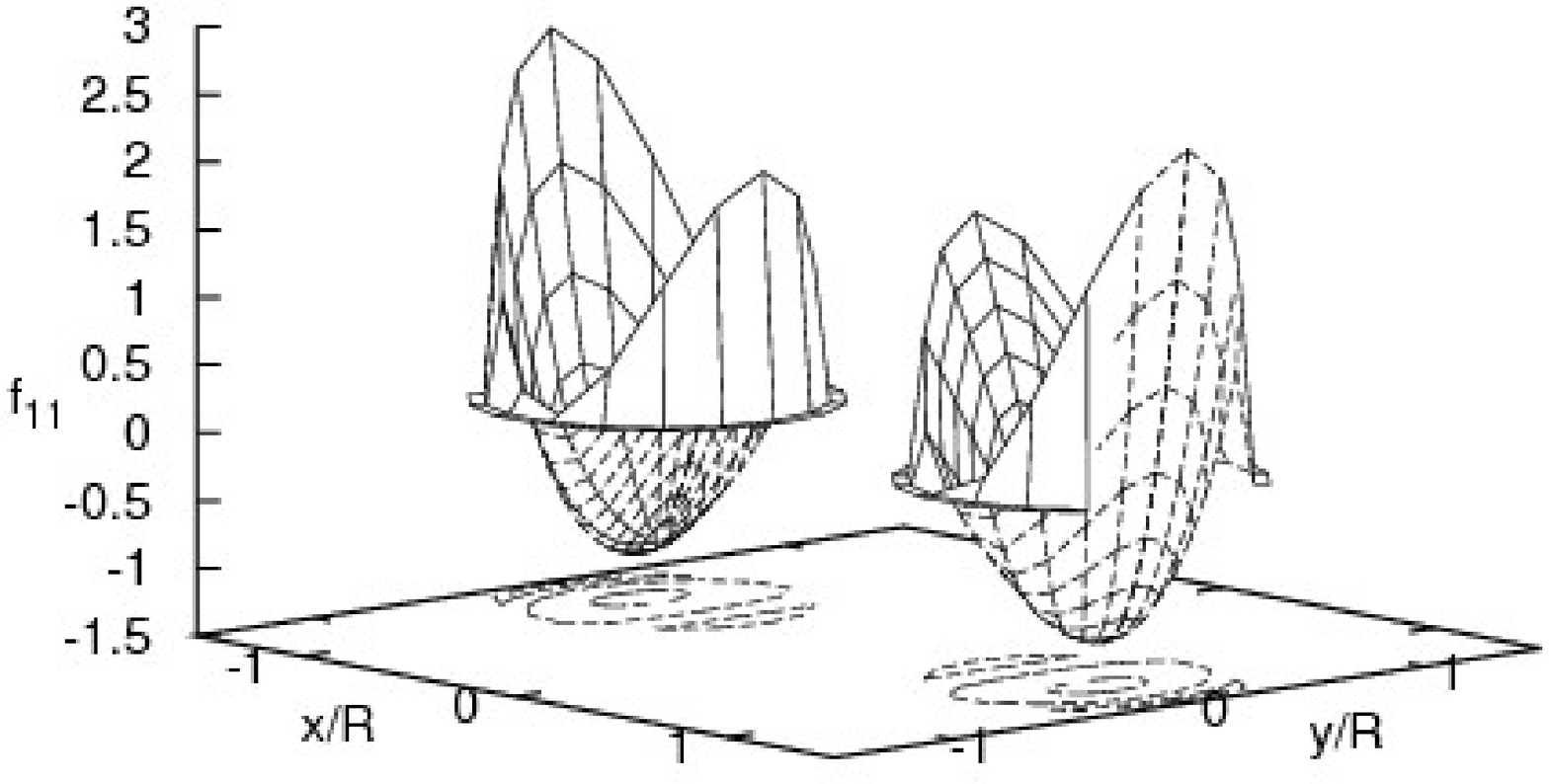}
\includegraphics[scale=0.42]{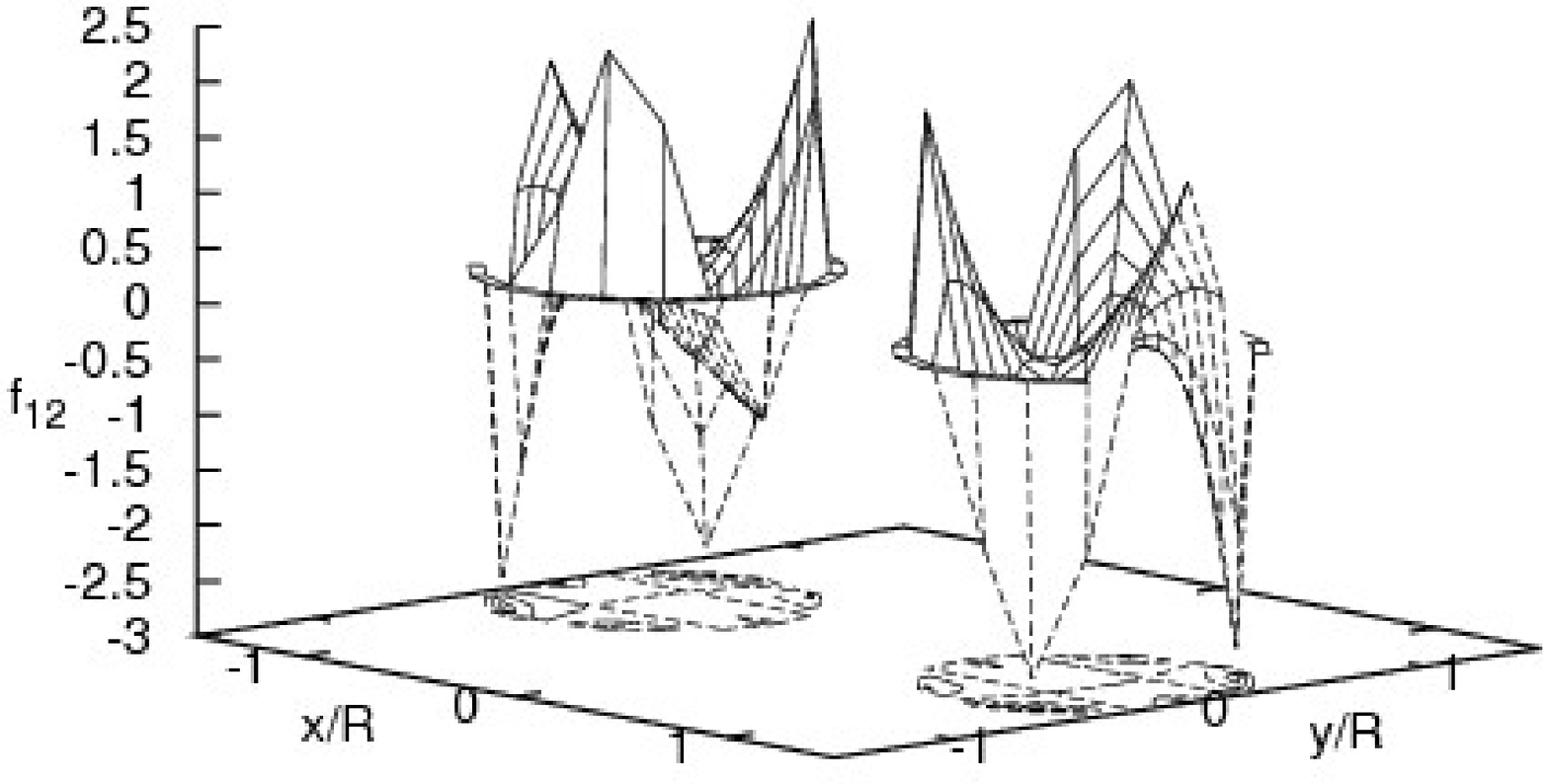}
\caption{The basis functions $f_7$ to $f_{12}$ for $q=1/2$.
}
\label{fig.displ9}
\end{figure}

\begin{figure}
\includegraphics[scale=0.42]{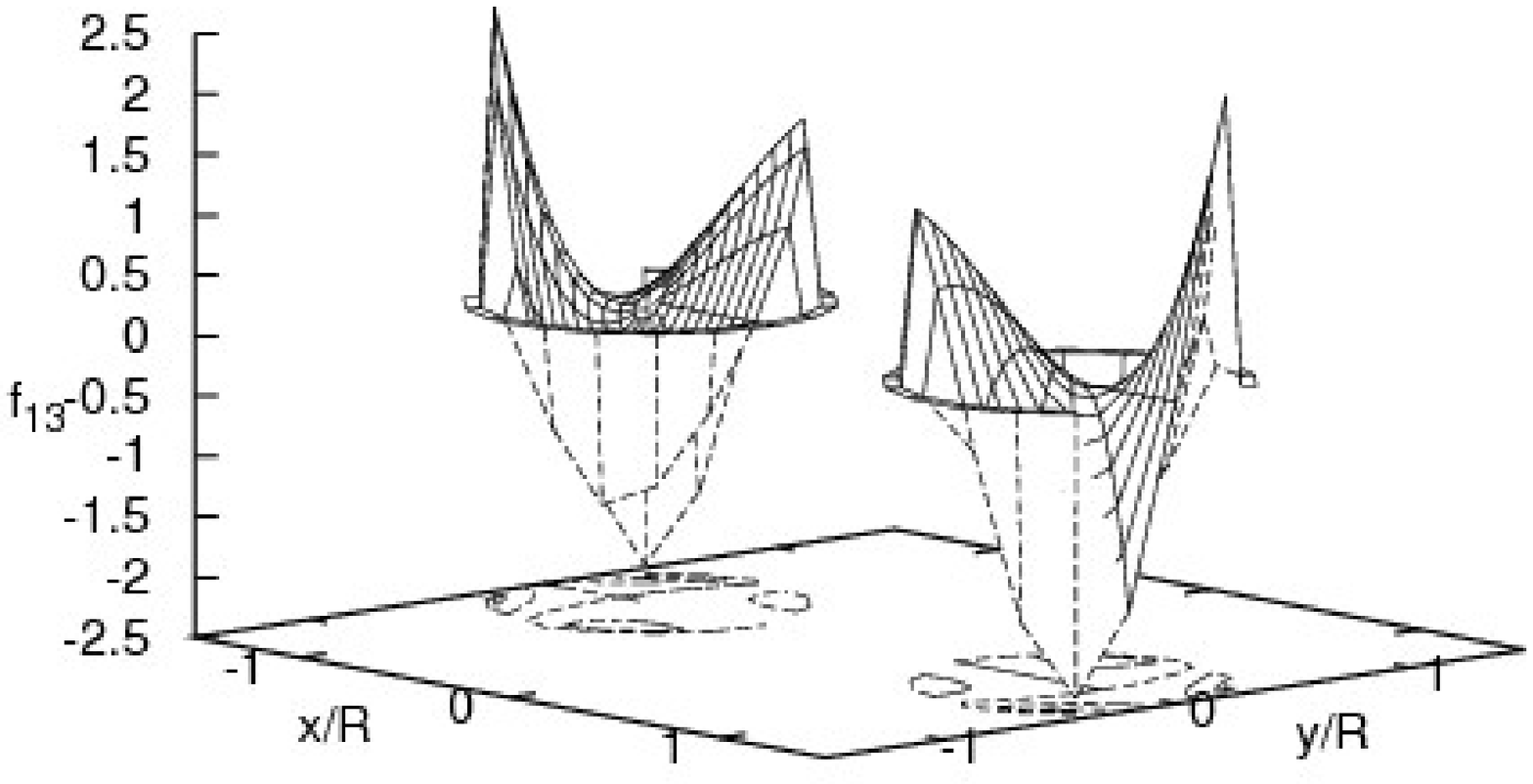}
\includegraphics[scale=0.42]{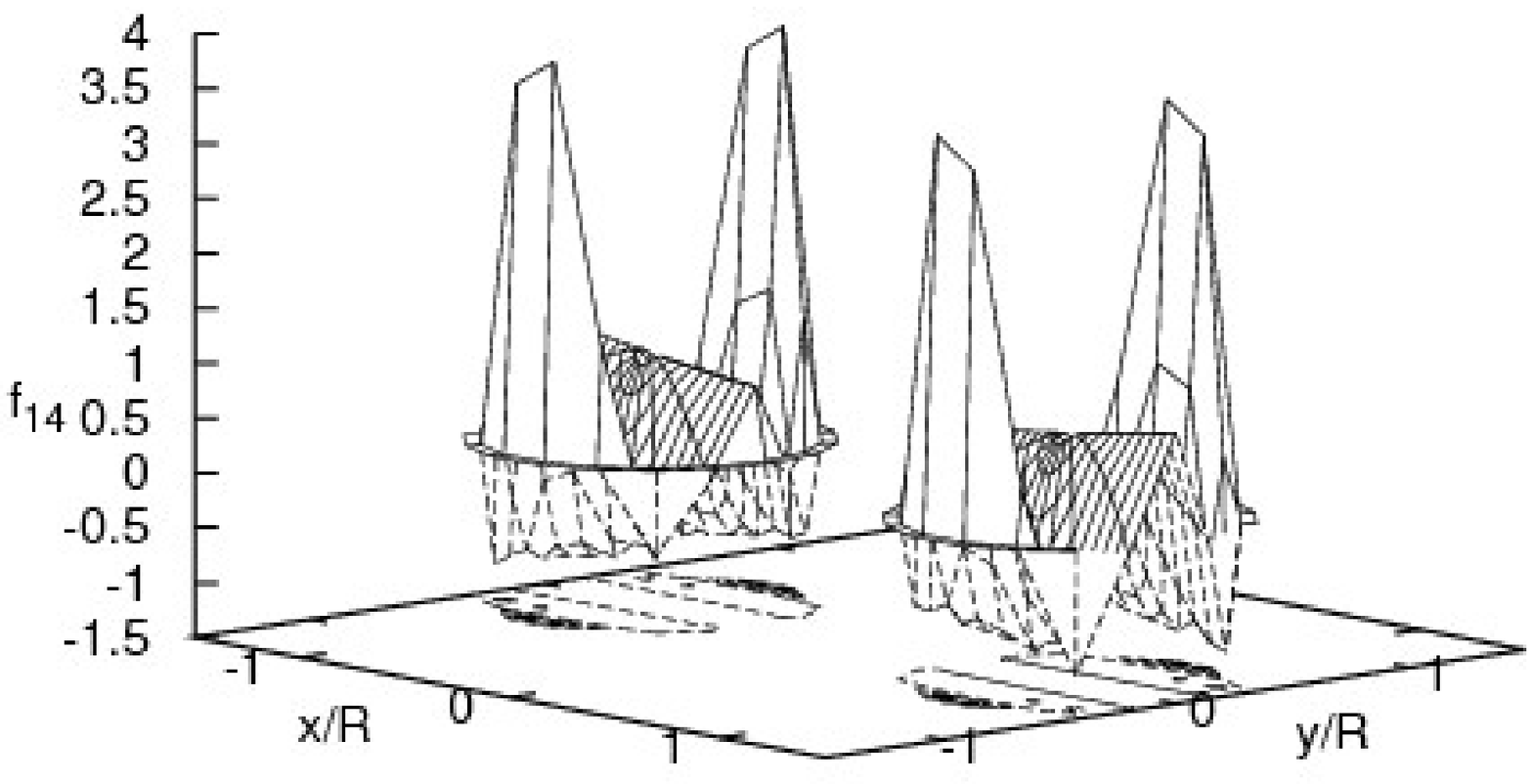}
\includegraphics[scale=0.42]{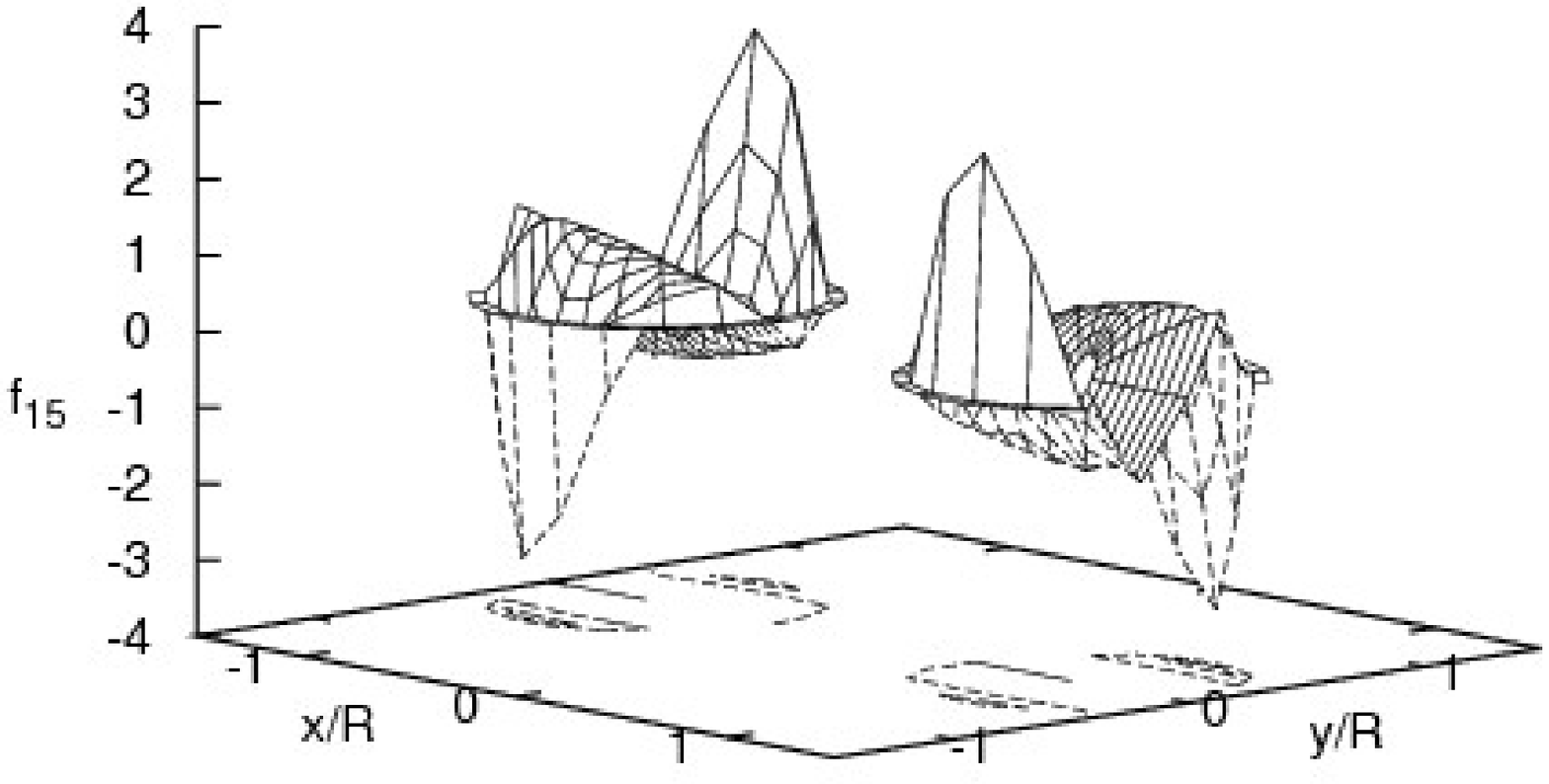}
\includegraphics[scale=0.42]{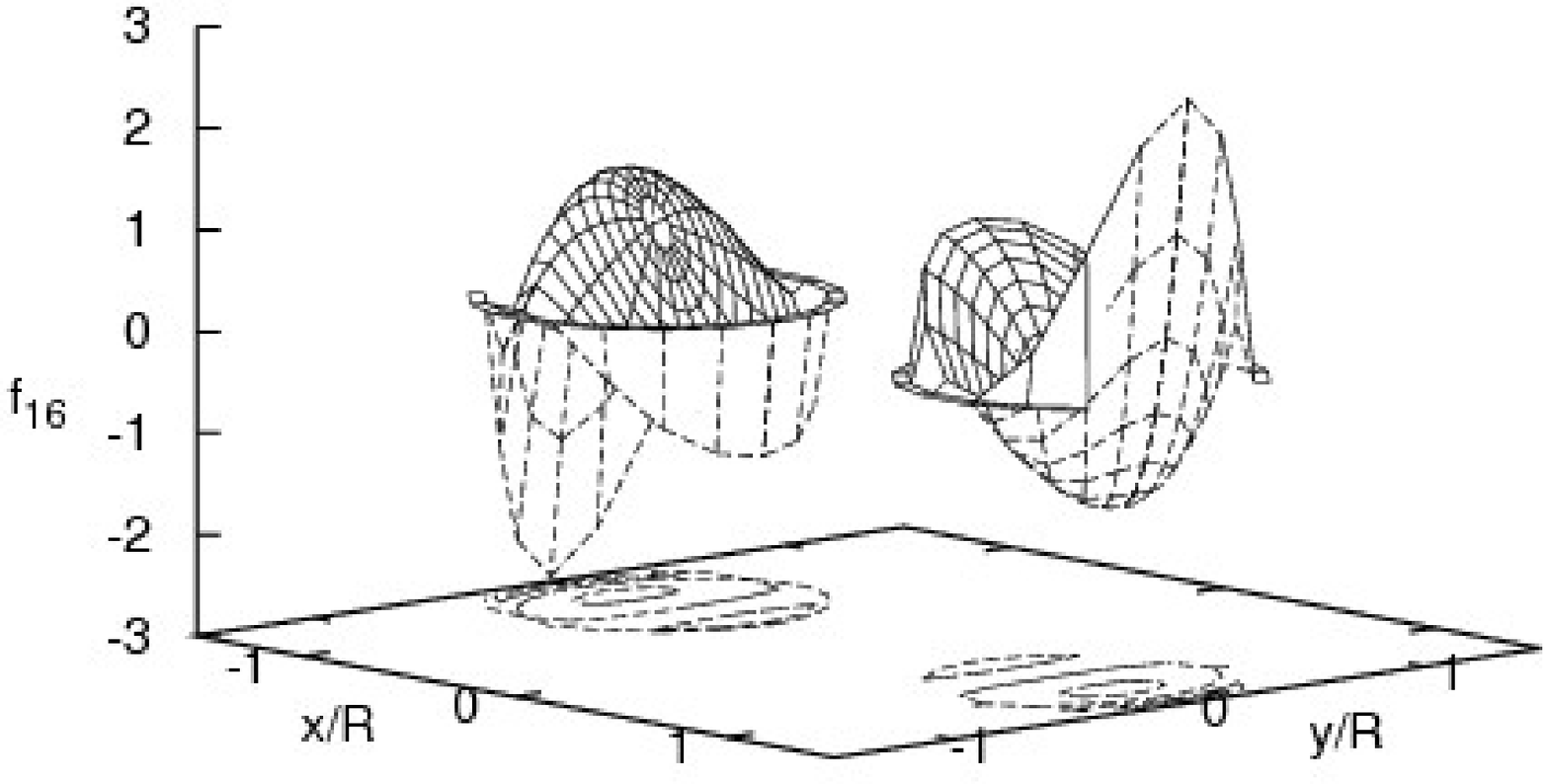}
\includegraphics[scale=0.42]{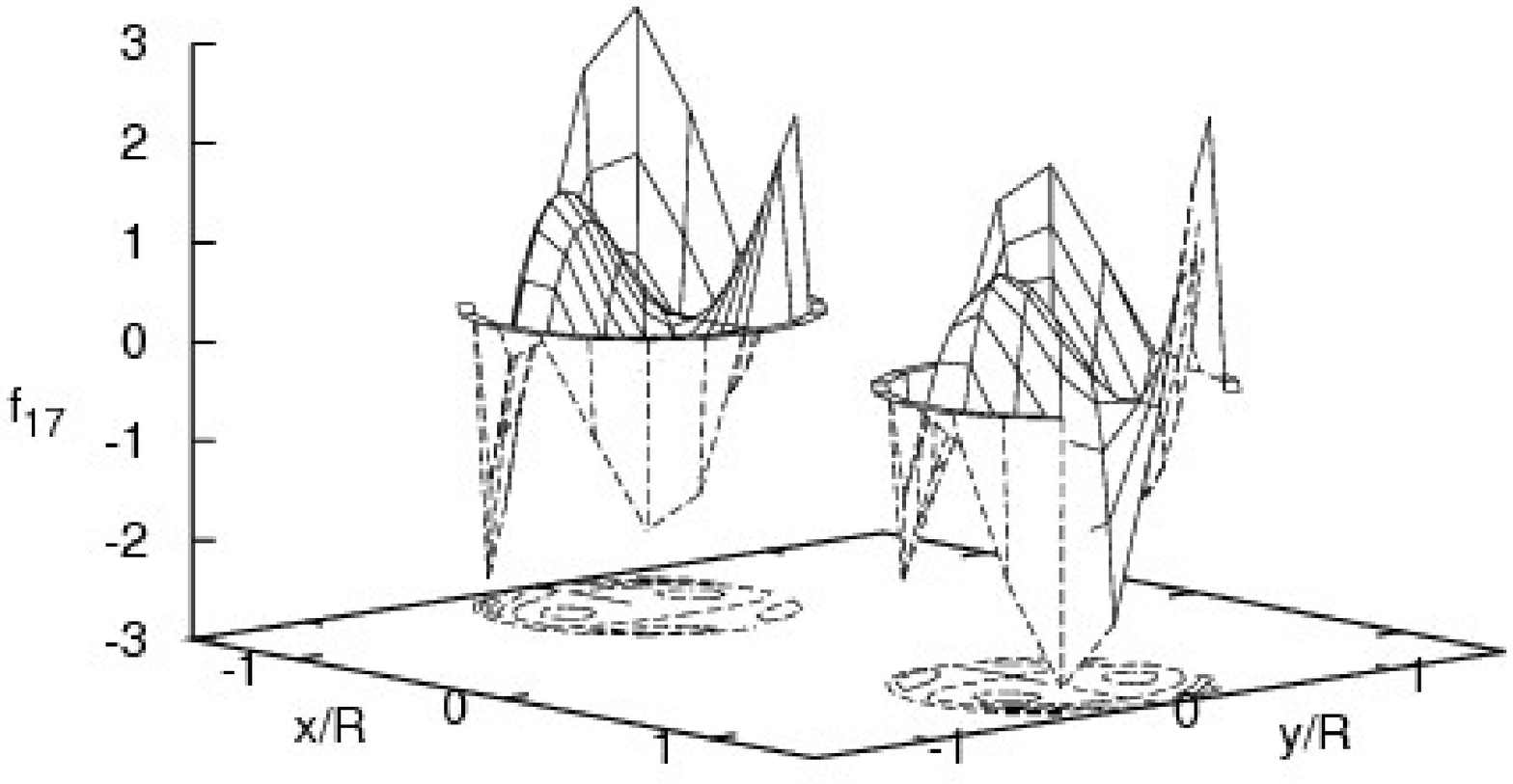}
\includegraphics[scale=0.42]{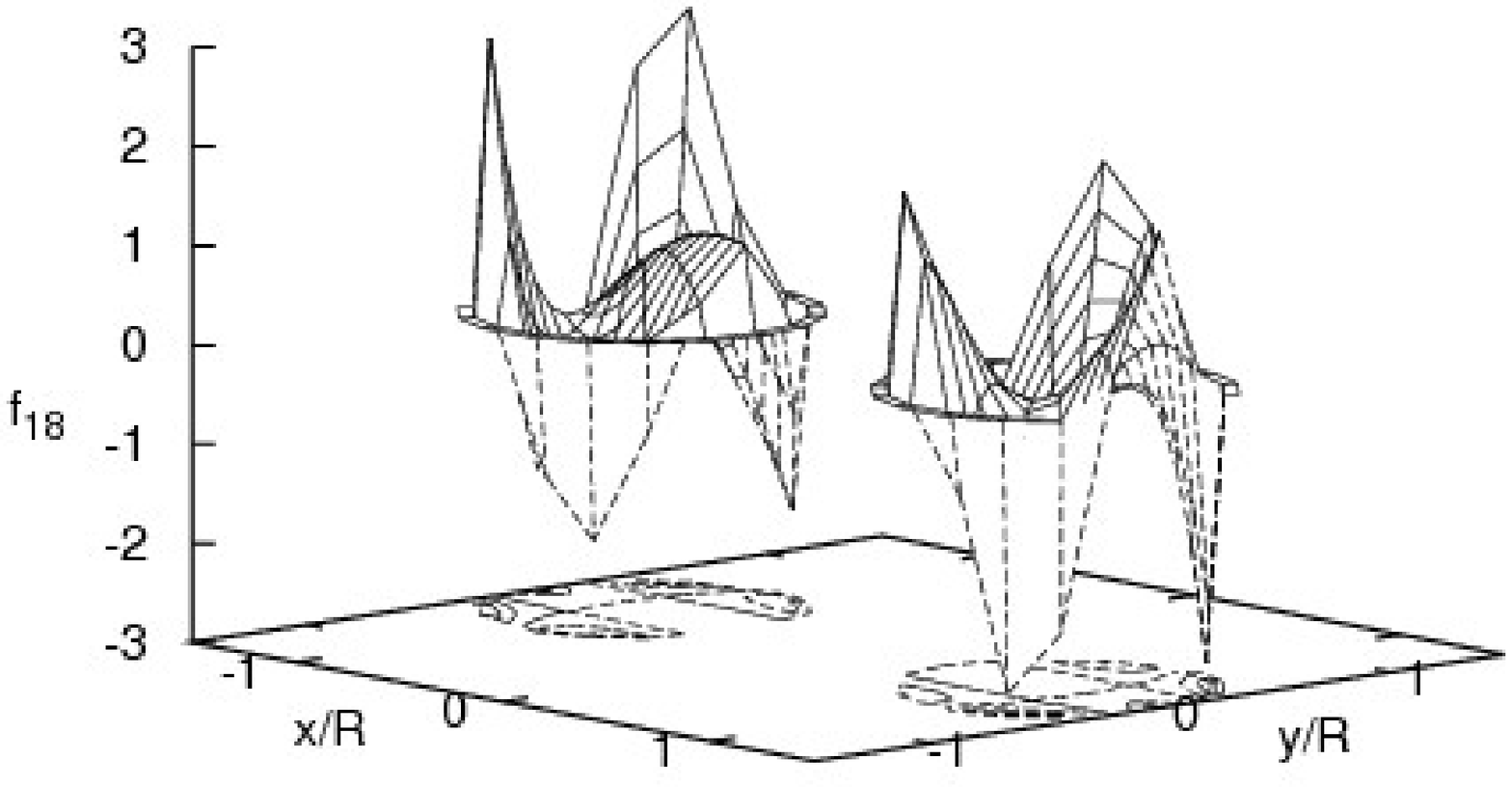}
\caption{The basis functions $f_{13}$ to $f_{18}$ for $q=1/2$.
}
\label{fig.displ13}
\end{figure}

Figs.\ \ref{fig.displ1}--\ref{fig.displ13} illustrate the first eighteen
of these functions for
$q=1/2$.

\section{Interferometric Signal} 
Expansions in orthogonal bases lead to accelerated book-keeping:
the integral over the square of
a function becomes the sum of the squared expansion coefficients (Parseval's equation).
In Maxwellian electrodynamics, the function is one of the two polarizations of the
electric field vector ${\cal E}$, and the simplification addresses how much
total energy passes through the cross-section.
If the circles are the entrance to a two-beam interferometer, this
addresses computation of the \emph{photometric} signal.

For a field expanded in the primitive basis,
\begin{equation}
{\cal E}(z,\theta)=\sum_k \eta_k z^{n_k}e^{im_k\theta},
\end{equation}
the \emph{interferometric} signal correlates
values at conjugated points $P=\pm 1$,
$x=P+s\cos\varphi$, $y=s\sin\varphi$,
sharing the
same local radial
coordinate $s$ and azimuth $\varphi$
(Fig.\ \ref{fig.geo}).
The coordinate transformation to
global $(z,\theta)$
circular coordinates are
\begin{equation}
z=\sqrt{x^2+y^2}=\sqrt{1+2Ps\cos\varphi+s^2};
\quad
\theta=\arctan(y/x)=(\pi+)\arctan\frac{s\sin\varphi}{P+\cos\varphi},
\label{eq.xy2sphi}
\end{equation}
where $(+\pi)$ indicates that $\pi$ is to be added for $P=-1$
if $\arctan$ denotes the principal value.
The interferometric signal (spatial autocorrelation)
is calculated by multiplying ${\cal E}$ at two conjugated points in the pupils,
a distance $2R$ apart in the $x$-direction, and integrating over
$s$ and $\varphi$. After transformation of ${\cal E}$ at $(z,\theta)$
to the individual $(s,\varphi)$ coordinates, the interferometric signal breaks
down into a sum over products of the expansion coefficients in terms of
these shifted/scaled polynomials.
As we have set up each $f_k$ as a linear combination of Zernike Polynomials
$Z_{j\le k}(r,\theta)$, the route to
transformations to polar coordinates originating at the two
circle's centers is known from the literature
\cite{CampbellJOSAA20,ComastriJOpt9,LundstromJOSAA24,SchwiegerlingJOSAA19,ShuJOSAA23}.
We summarize this in our notation;
this is off-topic in the sense that it is not related to
the orthogonality introduced above.

The interferometric intensity is
$\int {\cal E}^*_{P=-1}(s,\varphi){\cal E}_{P=+1}(s,\varphi)sdsd\varphi=\sum_{k,l}
\eta^*_k\eta_l I_{k,l}$, the generic information contained in
\begin{equation}
I_{k,l}\equiv \int_0^q s ds \int_0^{2\pi}d\varphi
(z^{n_k}e^{-im_k\theta})_{\mid{P=-1}}(z^{n_l}e^{im_l\theta})_{\mid{P=+1}}
\end{equation}
\begin{equation}
=
\int_0^q s ds\int_0^{2\pi}d\varphi
(1-2s\cos\varphi+s^2)^{n_k/2}
(1+2s\cos\varphi+s^2)^{n_l/2}
e^{-im_k[\pi+\arctan\frac{s\sin\varphi}{-1+s\cos\varphi}]}
e^{im_l \arctan\frac{s\sin\varphi}{1+s\cos\varphi}}
.
\end{equation}
We expand the integrand in a power series of $s$, integrate term by term
and show the results in form of the first terms of a power series in $q$.
The mean and excess of the four parameters,
\begin{equation}
m^{(+)}\equiv (m_k+m_l)/2;
\quad
m^{(-)}\equiv (m_k-m_l)/2;
\quad
n^{(+)}\equiv (n_k+n_l)/2;
\quad
n^{(-)}\equiv (n_k-n_l)/2
\end{equation}
are defined to
compress
the notation.
\begin{eqnarray}
(-1)^{m_k}I_{k,l}
&=&
\pi q^2
-\left[2m^{(+)}-n^{(-)}\right]^\dagger
\frac{\pi q^4}{8}
\nonumber \\
&&
-\left[(2m^{(+)}-n^{(-)})^2+2(2m^{(-)}-n^{(+)})\right] ^\dagger
\frac{\pi q^6}{192}
\nonumber \\
&&
+\left[2m^{(+)}-n^{(-)}\right] ^\dagger
\left[8+(2m^{(+)}-n^{(-)})^2+6(2m^{(-)}-n^{(+)})\right] ^\dagger
\frac{\pi q^8}{9216}
\nonumber \\
&&
+\Big[
(2m^{(+)}-n^{(-)})^4
+4(2m^{(+)}-n^{(-)})^2
\left\{8+3 (2m^{(-)}-n^{(+)})\right\}
 \nonumber \\
 && \quad
+12(2m^{(-)}-n^{(+)})
(4+2m^{(-)}-n^{(+)})
\Big] ^\dagger
\frac{\pi q^{10}}{737280}
\nonumber \\
&&
-\left[2m^{(+)}-n^{(-)}\right]^\dagger
\Big[
384+
(2m^{(+)}-n^{(-)})^4
+20(2m^{(+)}-n^{(-)})^2
(4+ 2m^{(-)}-n^{(+)})
 \nonumber \\
 && \quad
+20(2m^{(-)}-n^{(+)})
\left\{20+3(2m^{(-)}-n^{(+)})\right\}
\Big]^\dagger
\frac{\pi q^{12}}{88473600}
\nonumber \\
&&
-
\Big[
(2m^{(+)}-n^{(-)})^6
+10(2m^{(+)}-n^{(-)})^4
\left\{16+ 3(2m^{(-)}-n^{(+)})\right\}
 \nonumber \\
 && \quad
+4(2m^{(+)}-n^{(-)})^2
\left\{736+ 420(2m^{(-)}-n^{(+)})+45(2m^{(-)}-n^{(+)})^2\right\}
\nonumber \\
&& \quad
+120(2m^{(-)}-n^{(+)})
(8+2m^{(-)}-n^{(+)})
(4+2m^{(-)}-n^{(+)})
\Big]^\dagger
\frac{\pi q^{14}}{14863564800}
+\ldots
\end{eqnarray}
The symbol $\dagger$ indicates that bracket to its left represents
the product of two factors. The first factor is the bracketed term as written;
the second factor is the term
after the substitutions $n^{(-)}\rightarrow -n^{(-)}$ 
and $m^{(-)}\rightarrow -m^{(-)}$. So in the second factor of the product, some
sign flips occur whenever the total power of the $m^{(-)}$ and
$n^{(-)}$ is odd.

It is equivalent
to two re-expansions of the field at shifted centers of the sub-pupils
\cite{ComastriJOpt9} followed by areal integration.
Selection rules are implicit; if $2m^{(-)}-n^{(+)}$ or
$2m^{(+)}-n^{(-)}$ vanish, many coefficients in the power series become zero.

\section{Fourier Representation} 

All basis functions $f_j$ are linear superpositions of the primitive type
(\ref{eq.pBas})
after the $\cos(m\theta)$ are replaced by $[\exp(im\theta)+\exp(-im\theta)]/2$
and the
$\sin(m\theta)$ by $[\exp(im\theta)-\exp(-im\theta)]/(2i)$.
The two-dimensional Fourier Transform of these is
\begin{equation}
A_{n,m}(q,
\bm\sigma
)
\equiv \int \exp(2\pi i\bm{\sigma}\cdot \bm{z}) z^n \exp(im\theta)d\Omega
\end{equation}
for some wave number $\bm\sigma$.
These integrals are calculated
by individually translating each of the two circular areas to the origin of coordinates
as described in (\ref{eq.xy2sphi}),
\begin{equation}
A_{n,m}(q,\bm{\sigma})=
\sum_{P=\pm 1} e^{2\pi iP\sigma_x}
\int_{s\le q} e^{2\pi i \bm{\sigma}\cdot \bm{s}} z^ne^{im\theta}d^2s,
\label{eq.Asigma}
\end{equation}
where $\sigma_x$ is the component of $\bf\sigma$ along the baseline axis.
The coordinate transformation may interpret $ze^{i\theta}$ as
a complex variable. Since only
the cases of even $n-m$ need to be considered,
\begin{equation}
z^n e^{im\theta}
=
z^{n-m} (ze^{i\theta})^m
=
(x^2+y^2)^{(n-m)/2} (x+iy)^m
\end{equation}
is expanded in multinomials of $x$ and $y$, $x$ is replaced by $x+P$.
Finally the substitutions
\begin{equation}
x=s\cos\varphi=\frac{s}{2}\left(e^{i\varphi}+e^{-i\varphi}\right)
;
\quad
y=s\sin\varphi=\frac{s}{2i}\left(e^{i\varphi}-e^{-i\varphi}\right)
\end{equation}
express $z^n e^{im\theta}$ in the circular coordinates centered
at $x=P$ with radial coordinate $s$ and azimuth $\varphi$. Even powers of $P$
are dropped because $P^2=1$.
Table \ref{tab.znshift} demonstrates the cases for small $n$ and small
non-negative $m$.
\begin{table}
\begin{tabular}{p{0.5cm}p{0.5cm}p{16cm}}
$n$ & $m$ & $z^n \exp(im\theta)$ \\
\hline
0 & 0 & $1
$\\
1 & 1 & $P+s{e^{i\varphi}}
$\\
2 & 0 & $2\,Ps\cos \left( \varphi \right) +1+{s}^{2}
$\\
2 & 2 & $1+2\,Ps{e^{i\varphi}}+{s}^{2}{e^{2\,i\varphi}}
$\\
3 & 1 & $P+2\,P{s}^{2}+P{s}^{2}{e^{2\,i\varphi}}+2\,s{e^{i\varphi}}+s{e^{-i\varphi}}+{s}^
{3}{e^{i\varphi}}
$\\
3 & 3 & $P+3\,P{s}^{2}{e^{2\,i\varphi}}+3\,s{e^{i\varphi}}+{s}^{3}{e^{3\,i\varphi}}
$\\
4 & 0 & $2\,{s}^{2}\cos \left( 2\,\varphi \right) +2\, \left( 2\,P{s}^{3}+2\,Ps
 \right) \cos \left( \varphi \right) +1+4\,{s}^{2}+{s}^{4}
$\\
4 & 2 & $1+3\,P{s}^{3}{e^{i\varphi}}+P{s}^{3}{e^{3\,i\varphi}}+3\,Ps{e^{i\varphi}}+Ps{e^{
-i\varphi}}+3\,{s}^{2}+3\,{s}^{2}{e^{2\,i\varphi}}+{s}^{4}{e^{2\,i\varphi}}
$\\
4 & 4 & $1+4\,P{s}^{3}{e^{3\,i\varphi}}+4\,Ps{e^{i\varphi}}+6\,{s}^{2}{e^{2\,i\varphi}}+{
s}^{4}{e^{4\,i\varphi}}
$\\
5 & 1 & $P+2\,P{s}^{4}{e^{2\,i\varphi}}+6\,P{s}^{2}+6\,{s}^{3}{e^{i\varphi}}+3\,{s}^{3
}{e^{-i\varphi}}+2\,s{e^{-i\varphi}}+{s}^{5}{e^{i\varphi}}+P{s}^{2}{e^{-2\,i\varphi
}}+3\,P{s}^{4}+3\,P{s}^{2}{e^{2\,i\varphi}}+3\,s{e^{i\varphi}}+{s}^{3}{e^{3\,
i\varphi}}
$\\
5 & 3 & $P+4\,P{s}^{4}{e^{2\,i\varphi}}+4\,P{s}^{2}+6\,{s}^{3}{e^{i\varphi}}+s{e^{-i
\varphi}}+{s}^{5}{e^{3\,i\varphi}}+P{s}^{4}{e^{4\,i\varphi}}+6\,P{s}^{2}{e^{2\,i
\varphi}}+4\,s{e^{i\varphi}}+4\,{s}^{3}{e^{3\,i\varphi}}
$\\
5 & 5 & $P+{s}^{5}{e^{5\,i\varphi}}+5\,P{s}^{4}{e^{4\,i\varphi}}+10\,P{s}^{2}{e^{2\,i
\varphi}}+5\,s{e^{i\varphi}}+10\,{s}^{3}{e^{3\,i\varphi}}
$\\
6 & 0 & $2\,P{s}^{3}\cos \left( 3\,\varphi \right) +2\, \left( 3\,{s}^{2}+3\,{s}^{4
} \right) \cos \left( 2\,\varphi \right) +2\, \left( 3\,Ps+3\,P{s}^{5}+9\,
P{s}^{3} \right) \cos \left( \varphi \right) +1+9\,{s}^{4}+9\,{s}^{2}+{s}^
{6}
$\\
6 & 2 & $1+12\,P{s}^{3}{e^{i\varphi}}+4\,P{s}^{3}{e^{-i\varphi}}+4\,P{s}^{3}{e^{3\,i
\varphi}}+4\,Ps{e^{i\varphi}}+2\,Ps{e^{-i\varphi}}+8\,{s}^{2}+2\,P{s}^{5}{e^{3\,
i\varphi}}+{s}^{6}{e^{2\,i\varphi}}+4\,P{s}^{5}{e^{i\varphi}}+{s}^{2}{e^{-2\,i
\varphi}}+6\,{s}^{2}{e^{2\,i\varphi}}+8\,{s}^{4}{e^{2\,i\varphi}}+{s}^{4}{e^{4\,
i\varphi}}+6\,{s}^{4}
$\\
6 & 4 & $1+10\,P{s}^{3}{e^{i\varphi}}+10\,P{s}^{3}{e^{3\,i\varphi}}+5\,Ps{e^{i\varphi}}+P
s{e^{-i\varphi}}+5\,{s}^{2}+5\,P{s}^{5}{e^{3\,i\varphi}}+P{s}^{5}{e^{5\,i\varphi
}}+{s}^{6}{e^{4\,i\varphi}}+10\,{s}^{2}{e^{2\,i\varphi}}+10\,{s}^{4}{e^{2\,i
\varphi}}+5\,{s}^{4}{e^{4\,i\varphi}}
$\\
6 & 6 & $1+20\,P{s}^{3}{e^{3\,i\varphi}}+6\,Ps{e^{i\varphi}}+6\,P{s}^{5}{e^{5\,i\varphi}}
+{s}^{6}{e^{6\,i\varphi}}+15\,{s}^{2}{e^{2\,i\varphi}}+15\,{s}^{4}{e^{4\,i
\varphi}}
$\\
7 & 1 & $P+12\,P{s}^{4}{e^{2\,i\varphi}}+12\,P{s}^{2}+4\,P{s}^{4}{e^{-2\,i\varphi}}+18
\,{s}^{3}{e^{i\varphi}}+12\,{s}^{3}{e^{-i\varphi}}+{s}^{3}{e^{-3\,i\varphi}}+3\,
s{e^{-i\varphi}}+3\,{s}^{5}{e^{3\,i\varphi}}+12\,{s}^{5}{e^{i\varphi}}+6\,{s}^{5
}{e^{-i\varphi}}+P{s}^{4}{e^{4\,i\varphi}}+3\,P{s}^{2}{e^{-2\,i\varphi}}+18\,P{s
}^{4}+6\,P{s}^{2}{e^{2\,i\varphi}}+4\,s{e^{i\varphi}}+4\,{s}^{3}{e^{3\,i\varphi}
}+3\,P{s}^{6}{e^{2\,i\varphi}}+{s}^{7}{e^{i\varphi}}+4\,P{s}^{6}
$\\
7 & 3 & $P+20\,P{s}^{4}{e^{2\,i\varphi}}+10\,P{s}^{2}+2\,P{s}^{6}{e^{4\,i\varphi}}+20
\,{s}^{3}{e^{i\varphi}}+5\,{s}^{3}{e^{-i\varphi}}+2\,s{e^{-i\varphi}}+10\,{s}^{5
}{e^{3\,i\varphi}}+10\,{s}^{5}{e^{i\varphi}}+5\,P{s}^{4}{e^{4\,i\varphi}}+P{s}^{
2}{e^{-2\,i\varphi}}+10\,P{s}^{4}+10\,P{s}^{2}{e^{2\,i\varphi}}+{s}^{7}{e^{3
\,i\varphi}}+5\,s{e^{i\varphi}}+{s}^{5}{e^{5\,i\varphi}}+10\,{s}^{3}{e^{3\,i\varphi
}}+5\,P{s}^{6}{e^{2\,i\varphi}}
$\\
7 & 5 & $P+20\,P{s}^{4}{e^{2\,i\varphi}}+6\,P{s}^{2}+6\,P{s}^{6}{e^{4\,i\varphi}}+P{s}
^{6}{e^{6\,i\varphi}}+15\,{s}^{3}{e^{i\varphi}}+s{e^{-i\varphi}}+15\,{s}^{5}{e^{
3\,i\varphi}}+15\,P{s}^{4}{e^{4\,i\varphi}}+15\,P{s}^{2}{e^{2\,i\varphi}}+{s}^{7
}{e^{5\,i\varphi}}+6\,s{e^{i\varphi}}+6\,{s}^{5}{e^{5\,i\varphi}}+20\,{s}^{3}{e^
{3\,i\varphi}}
$\\
7 & 7 & $P+7\,P{s}^{6}{e^{6\,i\varphi}}+35\,P{s}^{4}{e^{4\,i\varphi}}+21\,P{s}^{2}{e^{
2\,i\varphi}}+{s}^{7}{e^{7\,i\varphi}}+7\,s{e^{i\varphi}}+21\,{s}^{5}{e^{5\,i
\varphi}}+35\,{s}^{3}{e^{3\,i\varphi}}
$\\
\end{tabular}
\caption{The primitive basis function in the integral (\ref{eq.Asigma}) translated
to $(s,\varphi)$ coordinates.
The cases of negative $m$ follow immediately
by complex-conjugation.
}
\label{tab.znshift}
\end{table}

This reduces each $A_{n,m}(q,\bm\sigma)$ to a finite sum of integrals
over centered
circles of radius $q$,
\begin{eqnarray}
\int_{s\le q} e^{2\pi i \bm{\sigma}\cdot \bm{s}} s^n e^{im\varphi}d^2s
&=&
e^{im\varphi_\sigma}\int_0^q s^{n+1}ds
\int_0^{2\pi} e^{2\pi i \sigma s \cos\varphi} e^{im\varphi}
d\varphi
\label{eq.Aprim}
\\
&=&
2\pi e^{im\varphi_\sigma} i^{|m|}
\int_0^q s^{n+1} J_{|m|}(2\pi s\sigma) ds
=
2\pi e^{im\varphi_\sigma} i^{|m|}
\frac{1}{(2\pi\sigma)^{n+2}} g_{n+1,|m|}(2\pi\sigma q)
,
\end{eqnarray}
where $\tan\varphi_\sigma\equiv \sigma_y/\sigma_x$ defines the
azimuth of the wave number. The
Bessel Function integrals are recursively computed via \cite[11.3.4]{AS}
\begin{equation}
g_{n+1,m}(\alpha)\equiv \int_0^\alpha t^{n+1}J_{m}(t)dt=\left\{
\begin{array}{ll}
\alpha^{n+1}J_{n+1}(\alpha)&, n=m;\\
\alpha^{n+1} J_{m+1}(\alpha)+(m-n) g(n,m+1,\alpha)&, n>m.
\end{array}
\right.
\end{equation}

The integrals $A_{n,m}(q)$ in (\ref{eq.Adef}) are just the special
case of zero momentum, $\sigma=0$. In this limit, (\ref{eq.Aprim}) simplifies to
$\int_{s\le q} s^n\exp(im\varphi)d^2s=2\pi\delta_{m,0}q^{n+2}/(n+2)$,
which proposes an alternative to compute Table \ref{tab.A}.

\begin{figure}
\includegraphics[scale=0.42]{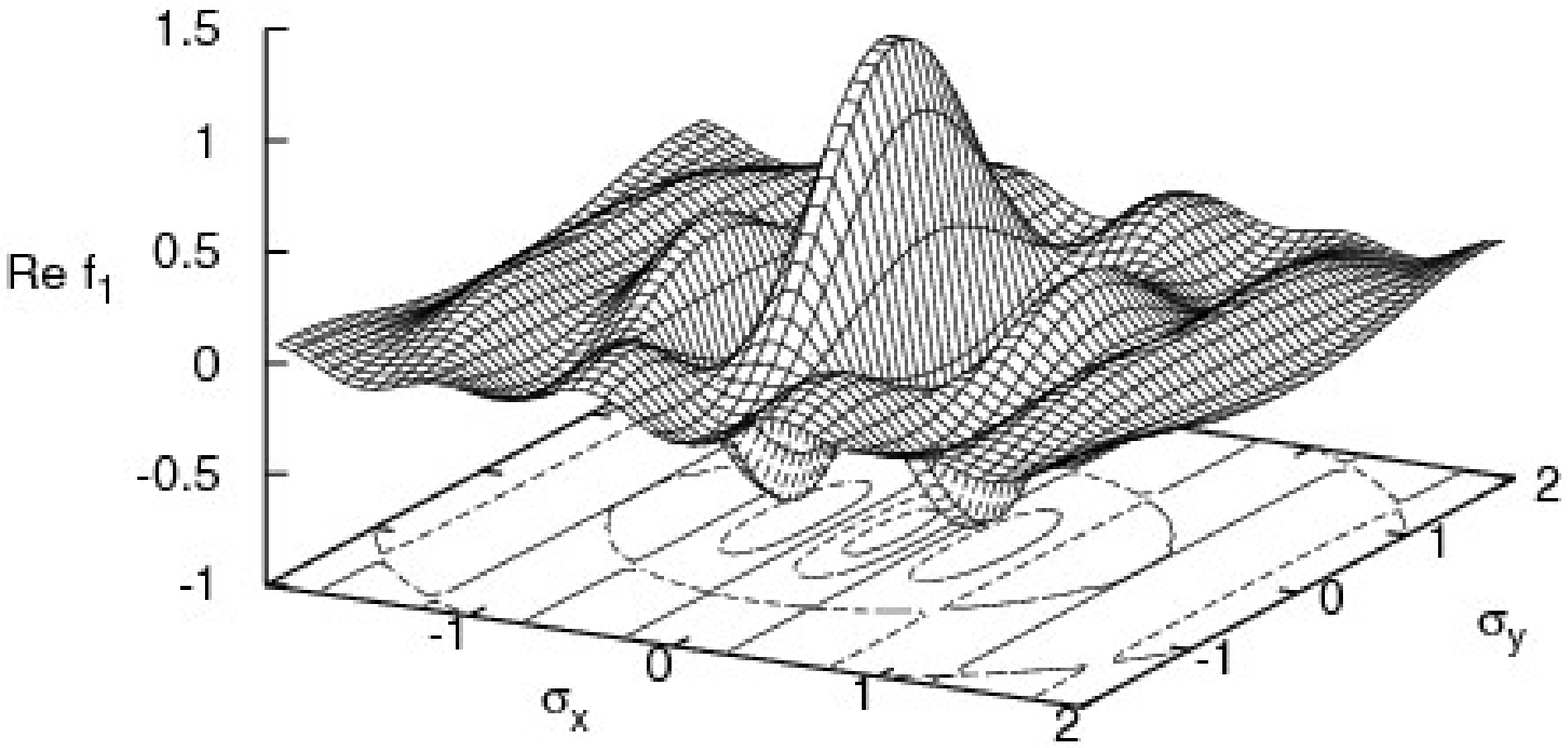}
\includegraphics[scale=0.42]{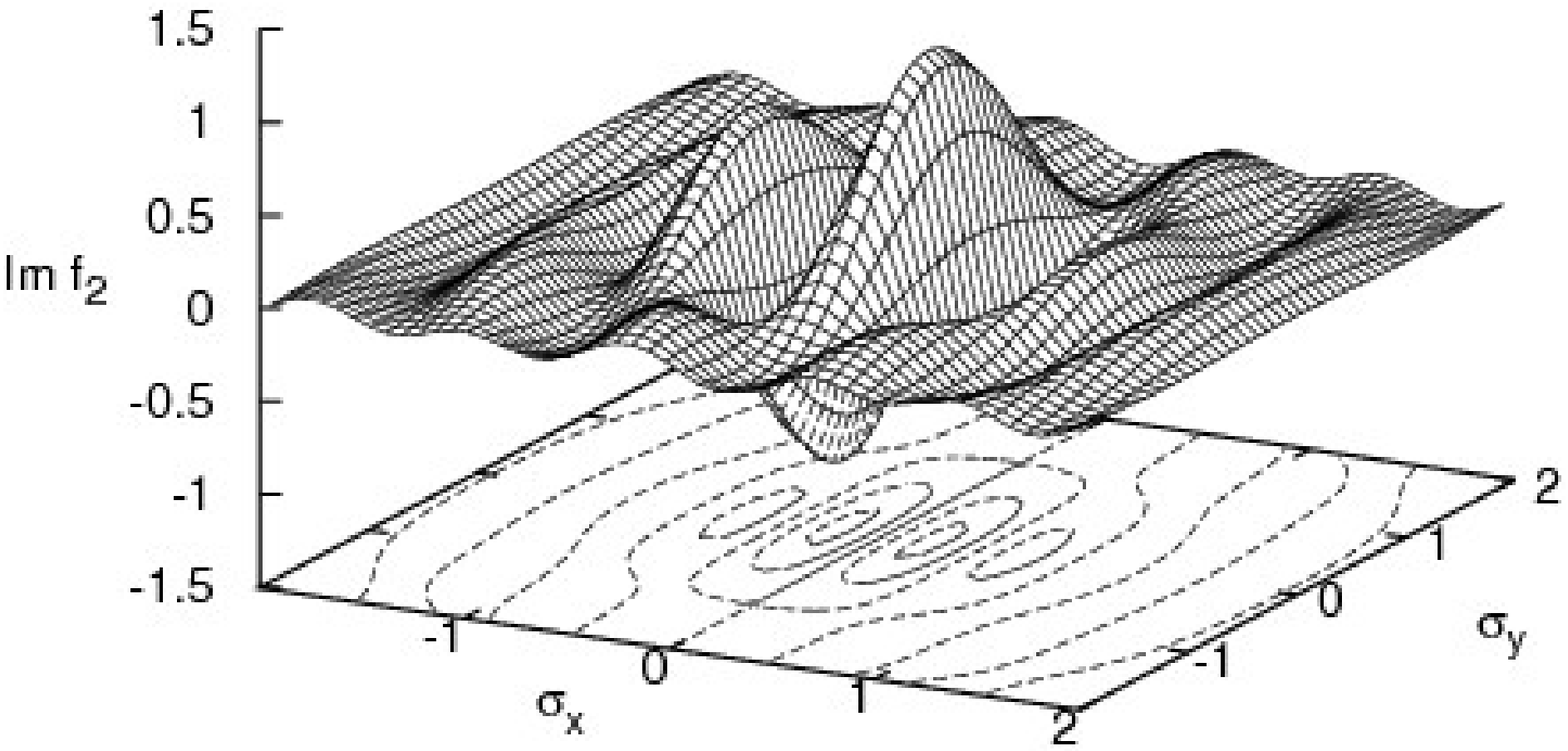}
\includegraphics[scale=0.42]{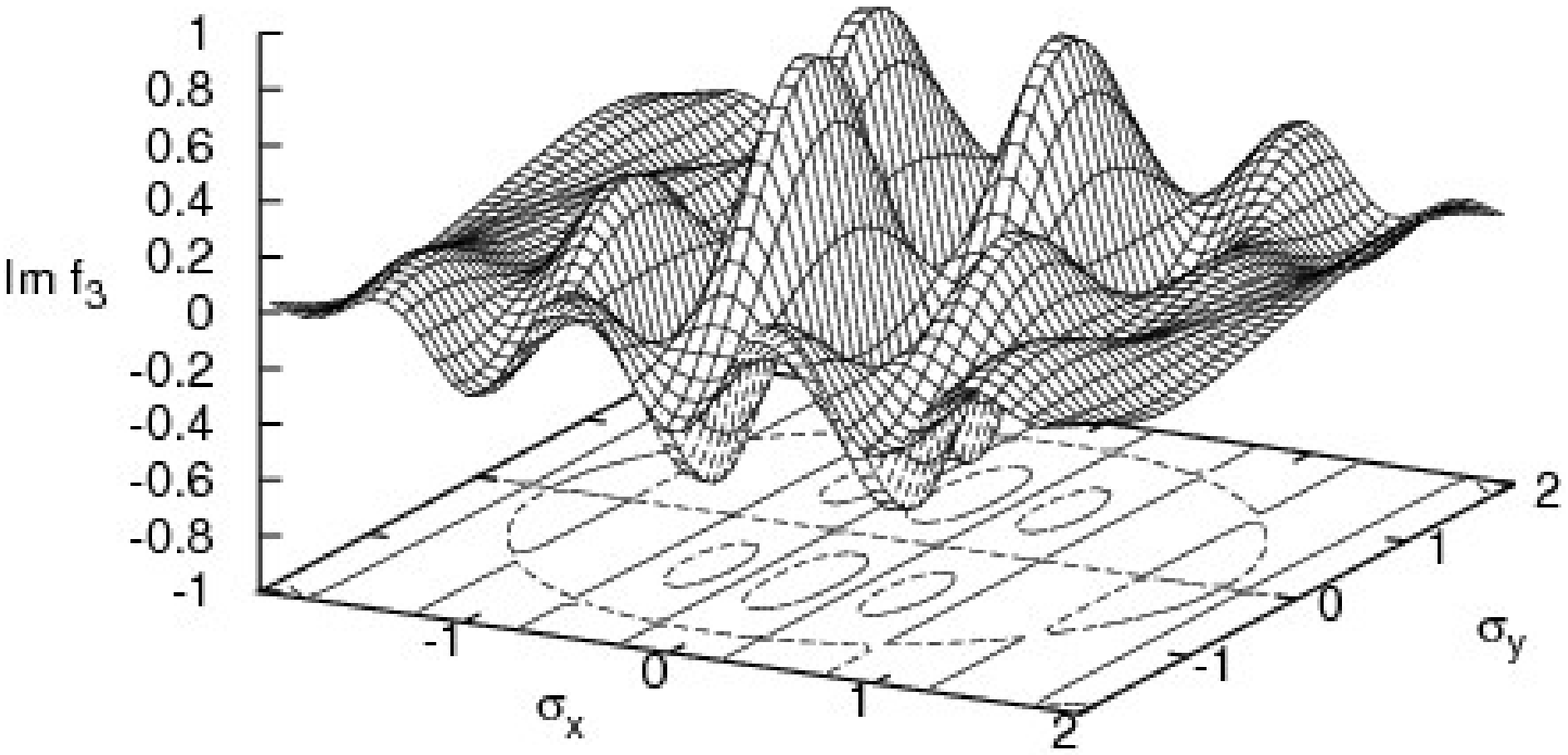}
\includegraphics[scale=0.42]{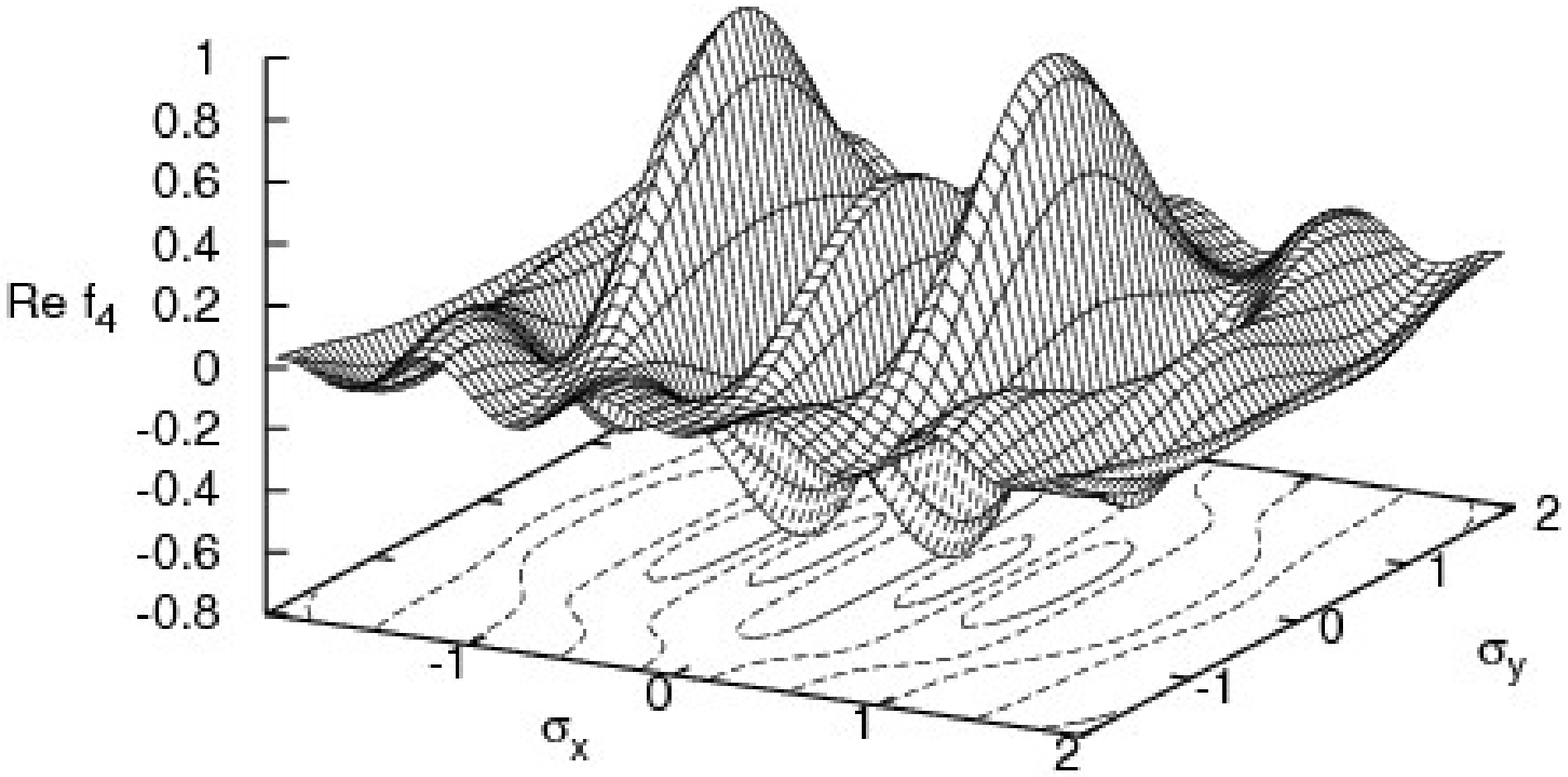}
\includegraphics[scale=0.42]{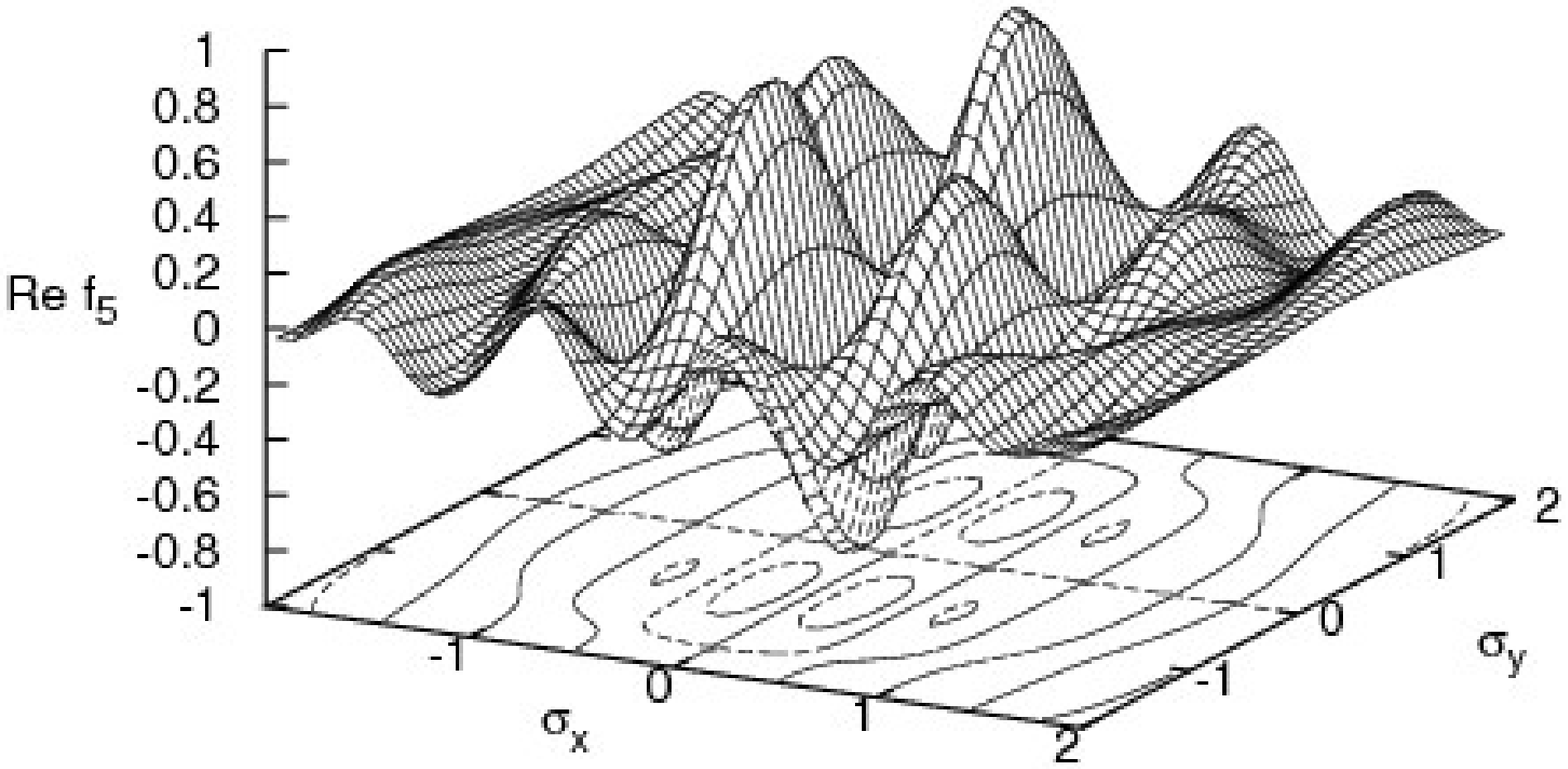}
\includegraphics[scale=0.42]{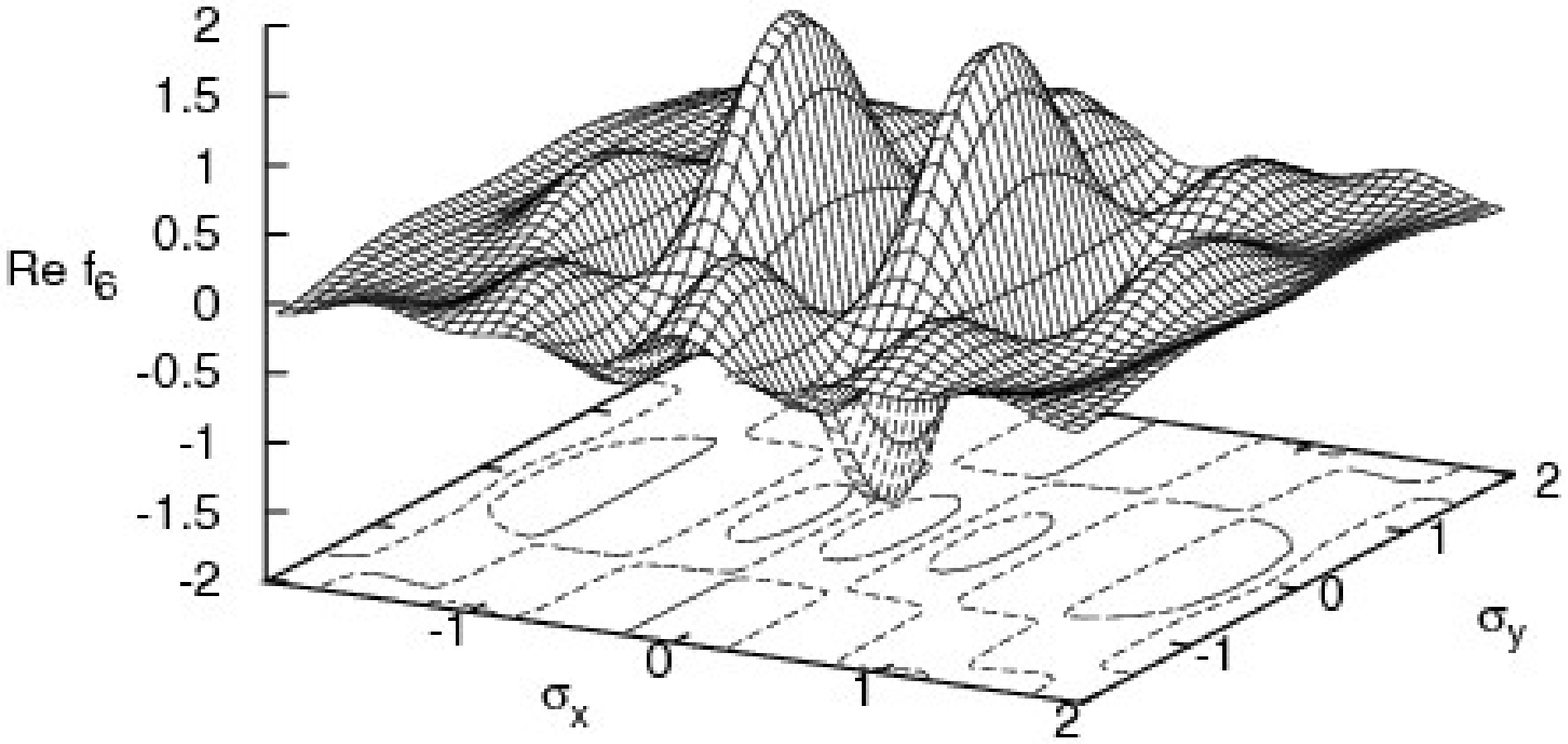}
\caption{The Fourier transforms of basis functions $f_1$ to $f_6$ for $q=1/2$.
}
\label{fig.displF1}
\end{figure}

\begin{figure}
\includegraphics[scale=0.42]{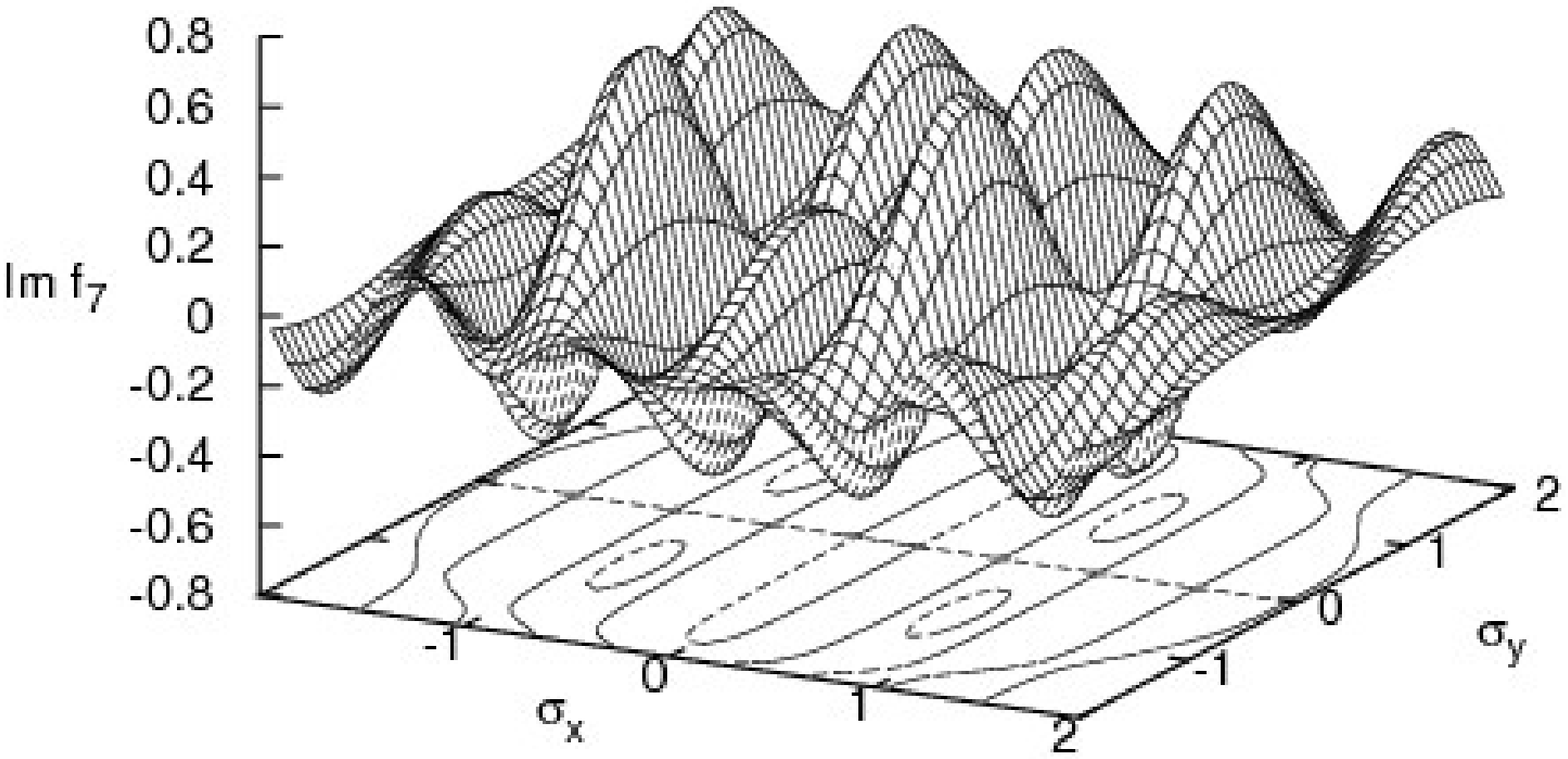}
\includegraphics[scale=0.42]{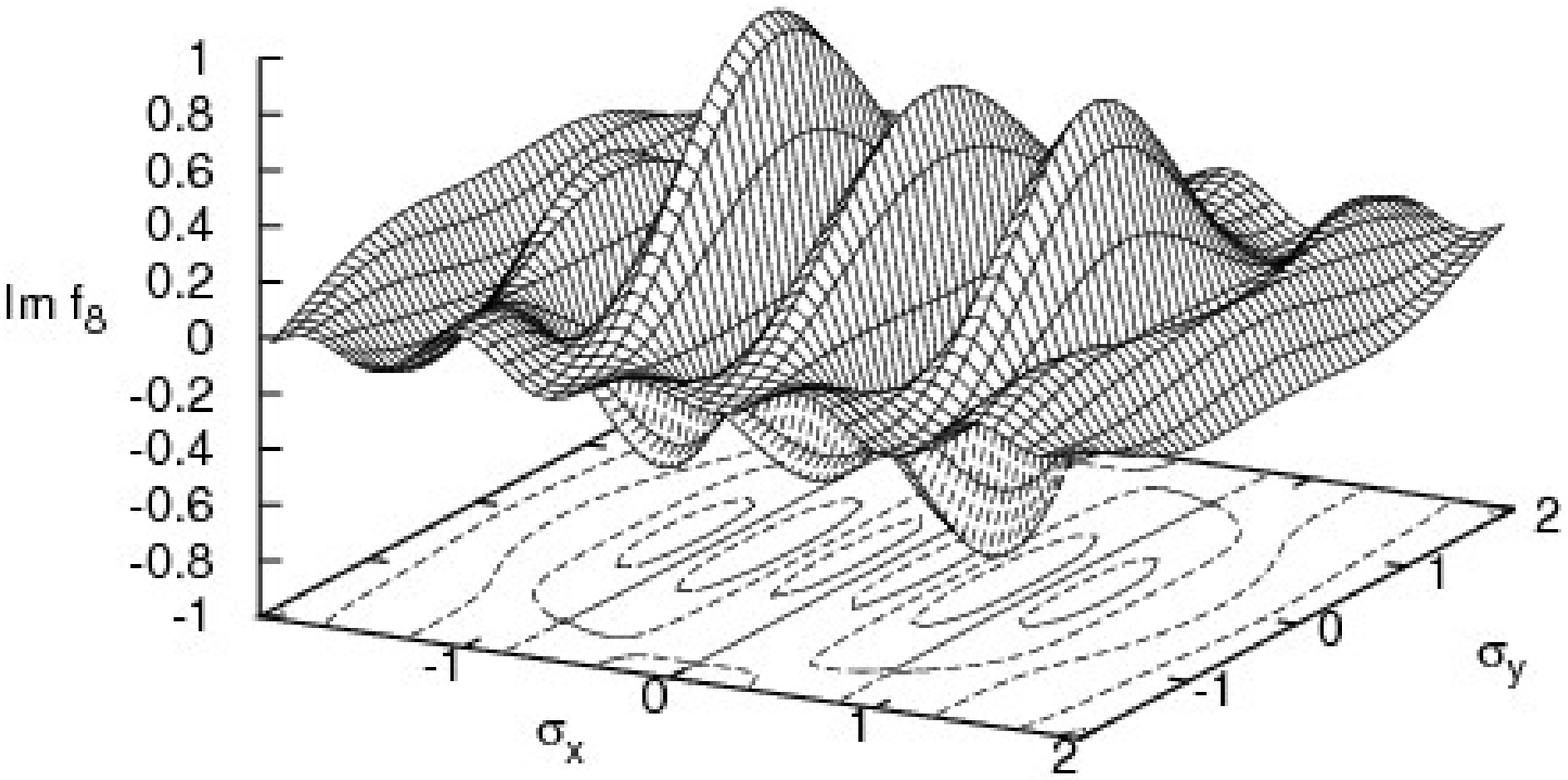}
\includegraphics[scale=0.42]{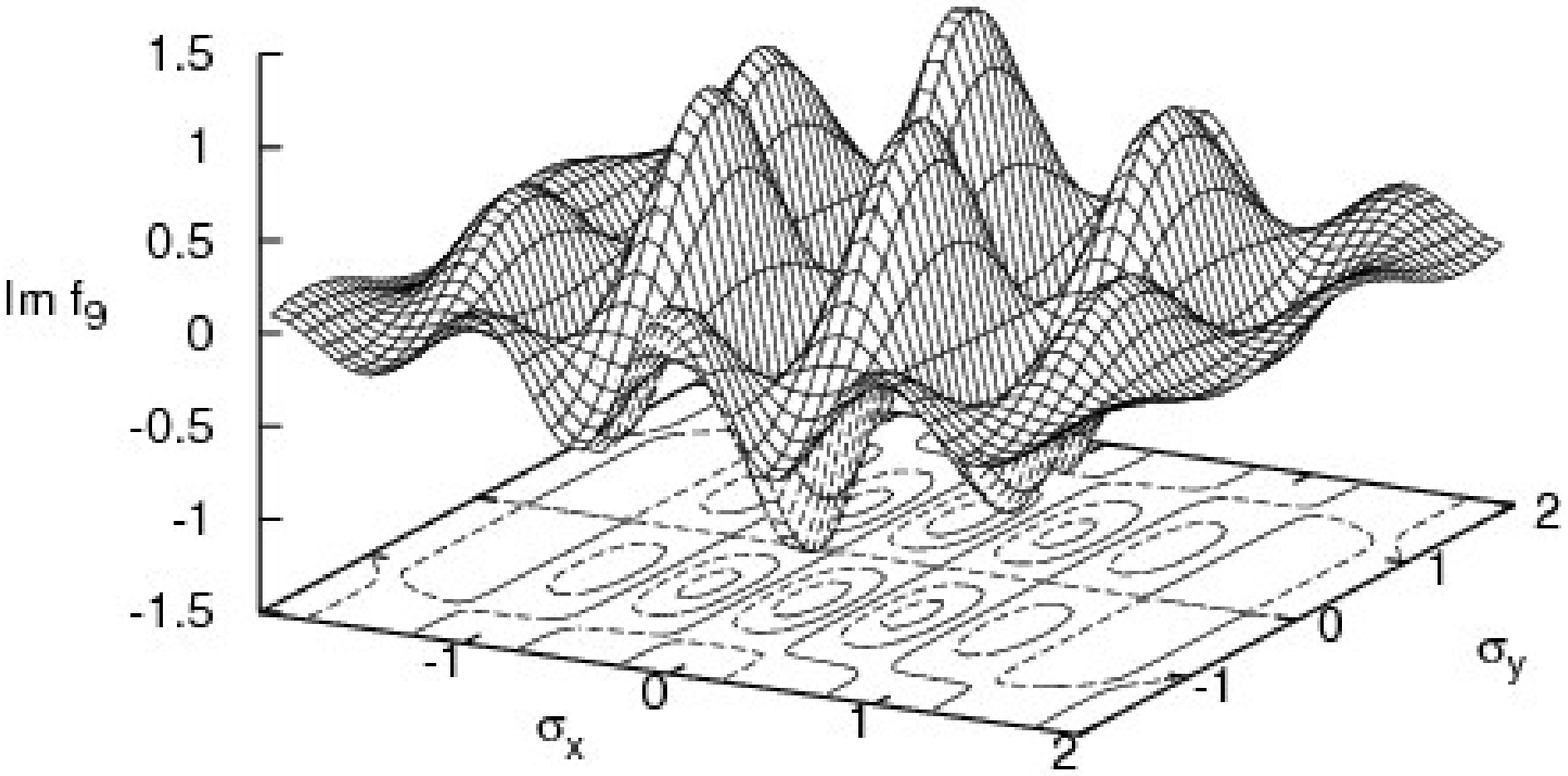}
\includegraphics[scale=0.42]{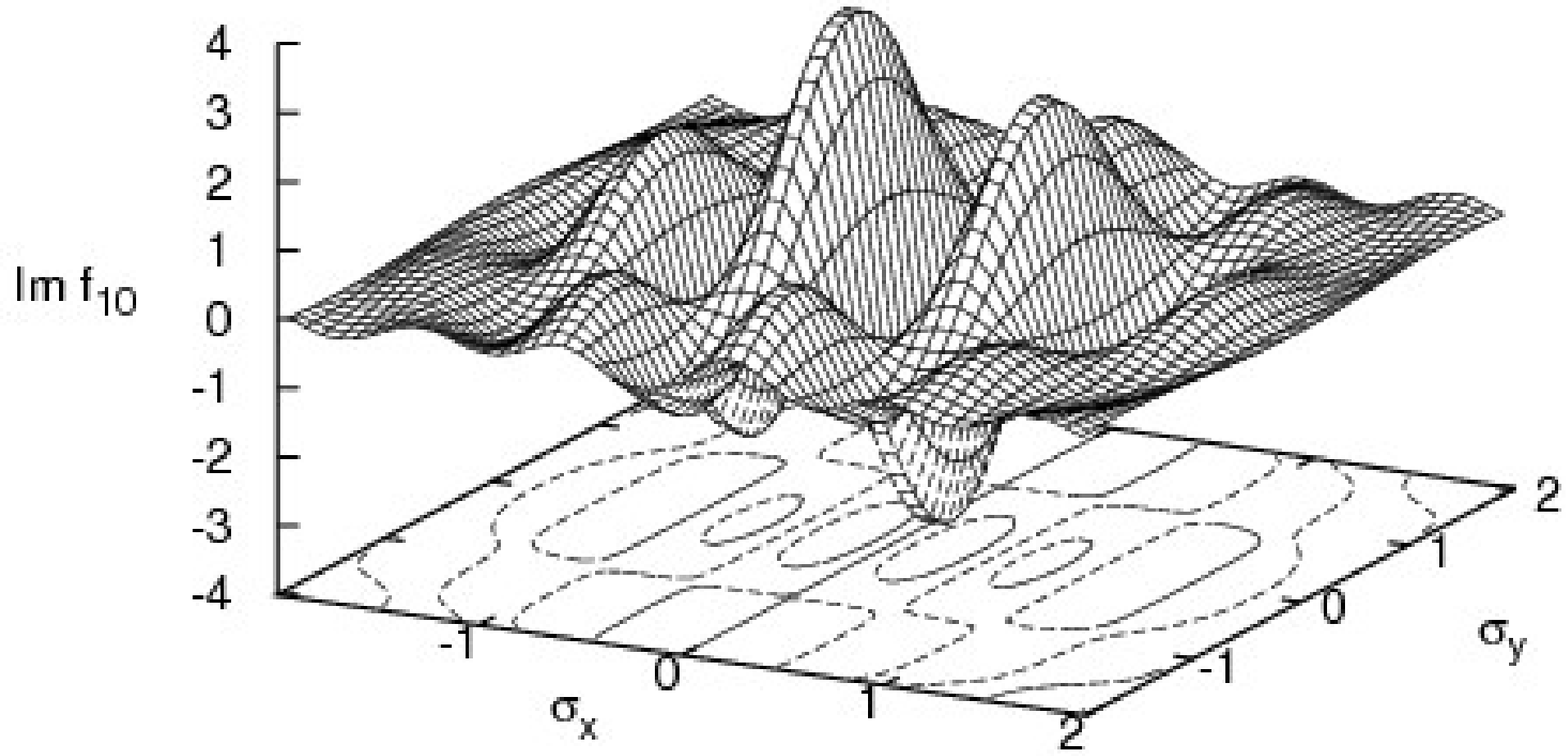}
\includegraphics[scale=0.42]{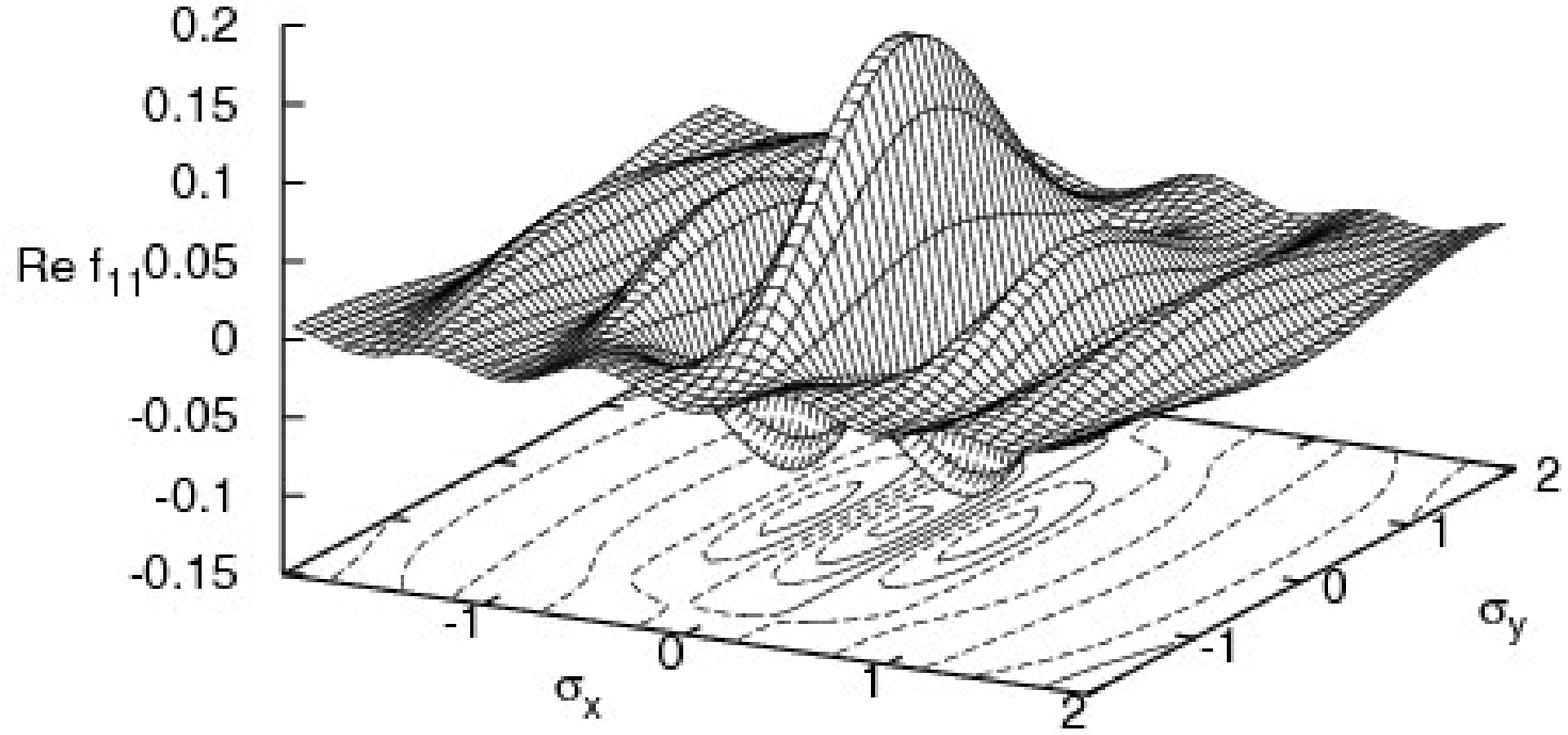}
\includegraphics[scale=0.42]{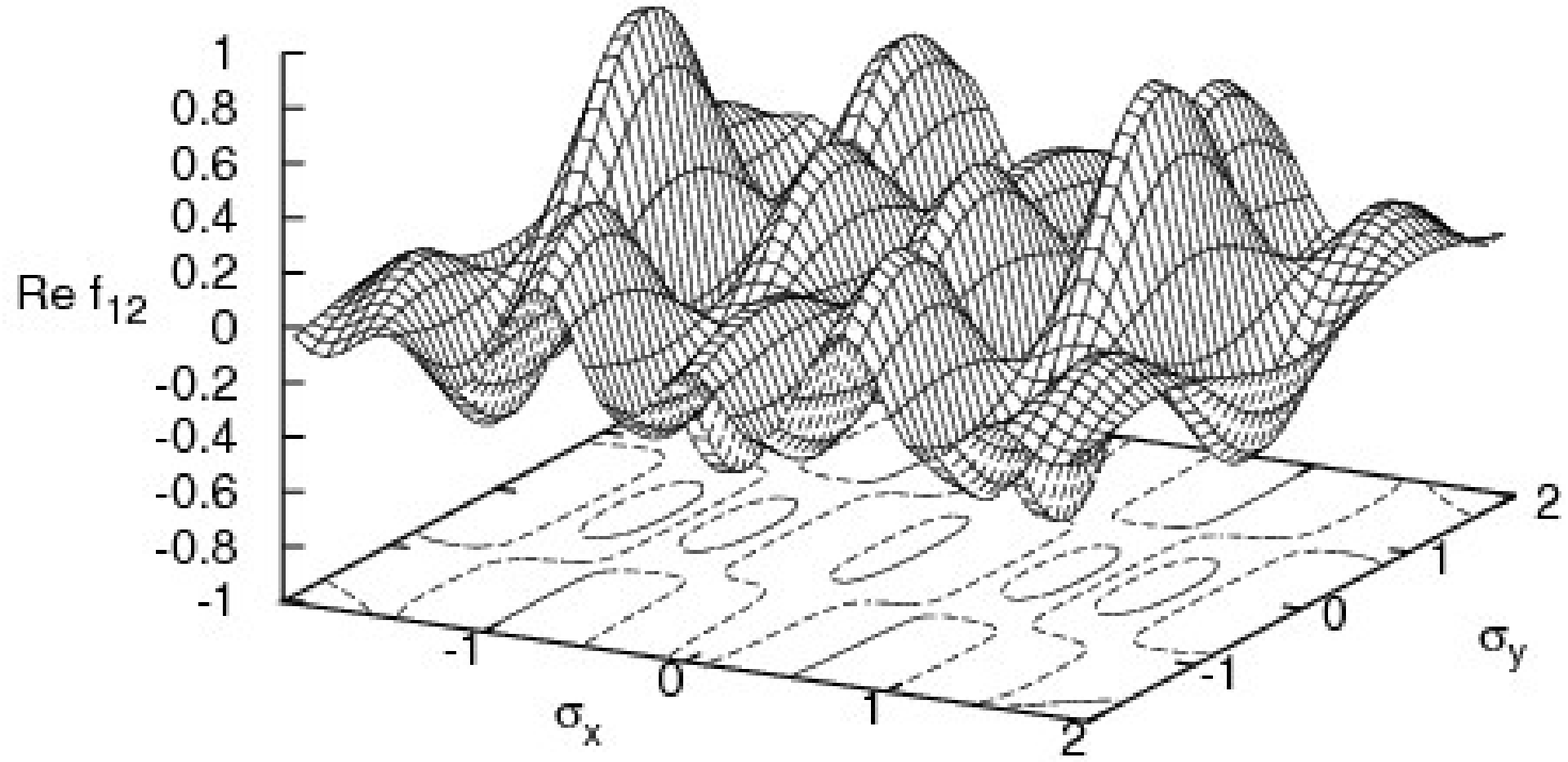}
\caption{The Fourier transforms of basis functions $f_7$ to $f_{12}$ for $q=1/2$.
}
\label{fig.displF7}
\end{figure}
We visualize the Fourier representations of the first 12 basis
functions in Figures \ref{fig.displF1} and \ref{fig.displF7}\@. The $f_j(x,y)$
have a well defined parity with respect to reflection across the origin, i.e.,
the terms in (\ref{eq.f1})--(\ref{eq.f13}) are superpositions of
$z^n\sin(m\theta)$ and $z^n\cos(m\theta)$ with a fixed, common parity $(-1)^m$
in each line.
As a consequence, the $f_j(\sigma_x,\sigma_y)$ are purely real or purely
imaginary, and only the non-vanishing of the two components is
shown.

$f_1$ in Figure \ref{fig.displF1} is the familiar diffraction limited
point spread function of the two-beam interferometer \cite{McCarthySPIE4006}.
$f_1$ in Figure \ref{fig.displ1}
is the \emph{convolution} of the double pinhole mask with a single circular
telescope pupil of area $\pi q^2$. Its Fourier representation
is the \emph{product} of the circular Airy disk centered at $\sigma=0$ by
the hyperbolic pattern of fringes with their narrower width along $\sigma_x$
as determined by the baseline length.

\section{Summary} 
Orthogonality of functions over the area of two non-overlapping
circles is defined according to the algebraic standards.
The fundamental areal integral for functions that
are of simple analytical format in the spherical coordinate system
attached to the center of symmetry has been recursively reduced
to polynomials or Gaussian hypergeometric functions of the normalized
circle radius. This defines a set of orthogonal basis functions
over the common area, if the Zernike basis functions, for example,
provide the input to the Gram-Schmidt procedure. The first few of these
have
been written down in analytical form. For the higher-order aberrations,
the complete information for numerical instantiation has been presented.

On a side note, computation of the interferometric signal of the pupil-beam
recombination given an expansion in the radial-azimuthal coordinates has
been reduced to a double sum of over a generic overlap integral.

\appendix
\section{Auxiliary integral $B$} 
\label{sec.B}
\subsection{Two Recursion Strategies.}

This section deals with the evaluation of the
integrals
defined in (\ref{eq.Bdef}). Splitting off and expanding
a square of the $(n+2)$nd power, we  obtain
\begin{equation}
B_{n+2,m}^\pm(q)
=
(q^2-1)B_{n,m}^\pm(q)
+2X_{n+1,m}^\pm(q)
,
\label{eq.Brecur}
\end{equation}
where we have introduced
\begin{equation}
X_{n+1,m}^\pm(q)
 \equiv 
\int_0^{\arcsin q}
\cos\theta \left[\cos\theta \pm\sqrt{q^2-\sin^2\theta}\right]^{n+1}\cos(m\theta)
d\theta
.
\label{eq.Xdef}
\end{equation}
This defines a first strategy to evaluate $B_{n+2,m}(q)$: recursive reduction
of the first lower index in steps of two at the expense of implementing
the $X_{n+1,m}(q)$ integrals---those to be treated in Appendix \ref{sec.X}\@.
Starting the recursion from an even $n$, one ends up at
\begin{equation}
B_{0,m}^\pm(q)
=
\frac{1}{m}
\sin(m\arcsin q)
\simeq 0.
\end{equation}
This is also correct in the limit $m\to 0$, and equivalent
to zero as the difference $B_{0,m}^+-B_{0,m}^-$ is formed.
The symbol $\simeq$ is reserved in this script to indicate that
terms
on the right hand side have been removed which cancel if differences
$B^+(q)-B^-(q)$ (common pair of subscripts with both $B$) or differences
$X^+(q)-X^-(q)$ (common pair of subscripts with both $X$) are calculated.
For odd $n$, the recursion (\ref{eq.Brecur}) terminates at
\begin{eqnarray}
B_{1,m}^\pm(q)
&=&
\int_0^{\arcsin q}d\theta \left(\cos\theta \pm \sqrt{q^2-\sin^2\theta}\right)\cos(m\theta)
\\
&\simeq&
\pm
\int_0^{\arcsin q}d\theta \sqrt{q^2-\sin^2\theta}T_m(\cos\theta).
\end{eqnarray}
For $m=0$ this value will be given in (\ref{eq.EK}).
For larger $m$, a mixture of partial integrations and the product rule
for Chebyshev polynomials \cite[22.7.24]{AS}
generates a recursion for the second index, again in steps of two \cite{NovarioETNA20}:
\begin{equation}
B_{1,m}^\pm(q)\simeq 
\frac{2(m-2)(1-2q^2)}{m+1}
B_{1,m-2}^\pm(q)
+\frac{5-m}{m+1}
B_{1,m-4}^\pm(q)
.
\label{eq.Bmrecur}
\end{equation}
Values to start this recursion at small even $m$ are discussed
in Section \ref{sec.Bodd}---this implies that $n+m$ is odd and does not
happen for the Gram-Schmidt seeds proposed in Section \ref{sec.Gschm}\@.
At small odd $m$ the recursion starts from
\begin{equation}
B_{1,1}^\pm(q)=\pm \frac{\pi}{4}q^2
; \qquad
B_{1,3}^\pm(q)=\pm \frac{\pi}{4}q^2(1-q^2)
.
\label{eq.B1113}
\end{equation}

An alternative second strategy to evaluate $B_{n+2,m}(q)$ looks as follows:
Binomial expansion of the $(n+2)$nd power in the
integrand of (\ref{eq.Bdef}) yields
\begin{multline}
B_{n+2,m}^\pm (q)\simeq \pm
\sum_{s=0}^{\lfloor(n+1)/2\rfloor}
\binom{n-s+1}{s}
(q^2-1)^s
\label{eq.Bnrecur}
\mathop{{\sum}'}_{\substack{u=0\\ n-u\; \text{odd}}}^{n+1-2s}
\binom{n+1-2s}{\frac{n+1-u}{2}-s}
\left[B_{1,u+m}^\pm(q)+B_{1,|u-m|}^\pm(q)\right]
.
\end{multline}
Here, the prime at the sum symbol means the term for $u=0$, if it occurs, is to be halved.
The differences to the first strategy are
\begin{itemize}
\item
No evaluation
of the $X_{n+1,m}^\pm(q)$ is needed. The contents of Appendix \ref{sec.X}
can be ignored.
\item
The decrement of the first index of $B_{n+2,m}^\pm$
via (\ref{eq.Bnrecur}) down to $1$ comes at the cost of an
increment of some second indices, eventually a recourse to (\ref{eq.Bmrecur}).
\item
Use of (\ref{eq.Bnrecur}) in conjunction with (\ref{eq.Bmrecur})
eventually uses (\ref{eq.B1113}),
whereas (\ref{eq.B1113}) and the case of odd $m$ are made irrelevant
for the first strategy through (\ref{eq.AofB}).
\end{itemize}
Explicit examples of (\ref{eq.Bnrecur}) for small $n$ are:
\begin{equation}
B_{2,m}^\pm \simeq
B_{1,m+1}+B_{1,m-1}
, \qquad m\ge 2;
\end{equation}
\begin{equation}
B_{3,m}^\pm \simeq
B_{1,m+2}+B_{1,m-2}+(1+q^2) B_{1,m}
, \qquad m\ge 2;
\end{equation}
\begin{equation}
B_{4,2}^\pm \simeq
2(1+q^2)B_{1,1}+(1+2q^2)B_{1,3}+B_{1,5} ;
\end{equation}
\begin{equation}
B_{4,m}^\pm \simeq
B_{1,m-3}
+(1+2q^2)B_{1,m-1}
+(1+2q^2)B_{1,m+1}
+B_{1,m+3}
, \quad m\ge 3;
\end{equation}
\begin{equation}
B_{5,2}^\pm \simeq
(1+3q^2)B_{1,0}+\left(2+4q^2+q^4\right)B_{1,2}
+(1+3q^2) B_{1,4}
+B_{1,6}
;
\end{equation}
\begin{equation}
B_{5,3}^\pm \simeq
\left(2+3q^2\right)B_{1,1}
+\left(1+4q^2+q^4\right)B_{1,3}
+(1+3q^2) B_{1,5}
+B_{1,7}
;
\end{equation}
\begin{eqnarray}
B_{5,m}^\pm
&\simeq &
B_{1,m-4}
+(1+3q^2)B_{1,m-2}
+\left(1+4q^2+q^4\right) B_{1,m}
\\
&&
+(1+3q^2) B_{1,m+2}
+ B_{1,m+4}
, \quad m\ge 4.
\nonumber
\end{eqnarray}
The argument $q$ at all the $B_{.,.}(q)$ has been omitted for brevity.

\subsection{The case of odd $n+m$.}
\label{sec.Bodd}
With a substitution $\sin\theta=\xi$,
$B_{1,0}$ can be written as a superposition of complete Elliptic
Integrals of the first and second kind:
\begin{eqnarray}
B_{1,0}^\pm(q)
&=&
\int_0^{\arccos\sqrt{1-q^2}}
(
\cos\theta \pm \sqrt{q^2-\sin^2\theta}
)
d\theta
\nonumber
\\
&=&
q\pm 
\int_0^q \frac{\sqrt{q^2-\xi^2}}{\sqrt{1-\xi^2}}d\xi
=
q\pm [E(q^2)-(1-q^2)K(q^2)]
.
\label{eq.EK}
\end{eqnarray}
\begin{equation}
B_{1,0}^\pm(q)\simeq \pm[E(q^2)-(1-q^2)K(q^2)]
.
\label{eq.B10q}
\end{equation}
A merger of the series expansions of the Elliptic Integrals yields
\begin{eqnarray}
B_{1,0}^\pm (q)
& \simeq &
\pm
\frac{\pi}{4}q^2\sum_{l=0}^\infty \left(\frac{q}{2}\right)^{2l}
\frac{[(2l-1)!!]^2}{l!(l+1)!}
=
\pm
\frac{\pi}{4} q^2
F\left(\begin{array}{cc}1/2,1/2 \\ 2\end{array}\mid q^2\right)
\nonumber
\\
&=&
\pm\frac{\pi}{4}q^2
\Big(
1+\frac{1}{8}q^2
+\frac{3}{64}q^4
+\frac{25}{1024}q^6
+\frac{245}{16384}q^8
+\frac{1323}{131072}q^{10}
\nonumber
\\
&&
+\frac{7623}{1048576}q^{12}
+\frac{184041}{33554432}q^{14}
+\frac{4601025}{1073741824}q^{16}
+\frac{29548805}{8589934592}q^{18}
+\frac{193947611}{68719476736}q^{20}
+\ldots
\Big)
\nonumber
\\
&
\xrightarrow[q\rightarrow 1]{}& \pm 1
.
\nonumber
\end{eqnarray}
Due to a logarithmic singularity at $q^2=1$, the power series converges poorly if the
argument $q^2$ of the Elliptic Integrals
approaches unity \cite{CodyMathComp19_249,LeeCPC60}.

(\ref{eq.B10q}) is the first anchor value for (\ref{eq.Bmrecur}).
The second is
\begin{eqnarray}
B_{1,2}^\pm(q)
&\simeq&
\pm\int_0^{\arcsin q}\sqrt{q^2-\sin^2q}\cos(2\theta) d\theta
\\
& \simeq &
\pm\frac{1}{3}(1-q^2)K(q^2)\pm \frac{1}{3}(2q^2-1)E(q^2)
.
\label{eq.B12q}
\end{eqnarray}
With the aid of (\ref{eq.Bmrecur}), expressions for $B_{1,m}^\pm(q)$ are
bootstrapped from (\ref{eq.B10q}) and (\ref{eq.B12q}). This sequence starts:
\begin{equation}
B_{1,4}^\pm(q) \simeq \pm\frac{1}{15}(q^2-1)(8q^2-1)K(q^2)
 \pm \frac{1}{15}(-16q^4+16q^2-1)E(q^2);
\end{equation}
\begin{equation}
B_{1,6}^\pm(q) \simeq \mp \frac{1}{105}(q^2-1)(128q^4-80q^2+3)K(q^2)
 \pm \frac{1}{105}(2q^2-1)(128q^4-128q^2+3)E(q^2);
\end{equation}
\begin{equation}
B_{1,8}^\pm(q) \simeq \pm\frac{1}{315}(q^2-1)(1024q^6-1152q^4+288q^2-5)K(q^2)
 \pm \frac{1}{315}(-2048q^8+4096q^6-2496q^4+448q^2-5)E(q^2).
\end{equation}

\section{Auxiliary integral $X$} 
\label{sec.X}

The auxiliary integrals (\ref{eq.Xdef}) are put into 
an algebraic format by the substitution $\sin\theta \equiv qz$,
\begin{equation}
X_{n+1,m}^\pm(q)
=
q\int_0^1\left(\sqrt{1-q^2z^2}\pm q\sqrt{1-z^2}\right)^{n+1}T_m(\sqrt{1-q^2z^2})dz,
\end{equation}
and then broken down through
binomial expansion of the $(n+1)$st power and the explicit
polynomial expression for the Chebyshev function \cite[22.3.6]{AS}
via \cite[3.197.3]{GR}
\begin{multline}
X_{n+1,m}^\pm (q)
\simeq
\pm
2^{m-3}\pi (n+1)! q^2
\sum_{j=0}^{\lfloor n/2\rfloor}
\frac{1}{(n-2j)!j!}
\left(\frac{q}{2}\right)^{2j}
\nonumber
\\
\times
\left\{
\begin{array}{ll}
\sum_{\sigma=0}^{\lfloor m/2\rfloor}(-\frac{1}{4})^\sigma
\frac{m}{2(m-\sigma)}
\binom{m-\sigma}{\sigma}
F\left(
\begin{array}{cc}
j+\sigma-\frac{n+m}{2},\frac{1}{2} \\
j+2
\end{array}
\mid
q^2
\right) & ; m>0;
\\
F\left(
\begin{array}{cc}
j-\frac{n}{2},\frac{1}{2} \\
j+2
\end{array}
\mid
q^2
\right) & ; m=0.
\end{array}
\right.
\nonumber
\end{multline}
If $n+m$ is an even number---which is the case for the Gram-Schmidt procedure
described in Section \ref{sec.f}---the hypergeometric series terminate
and become polynomials of $q^2$ of order $(n+m)/2-j-\sigma$ \cite[15.4.1]{AS}.
The result can be tabulated in terms of power series coefficients $\alpha_j(n,m)$,
\begin{equation}
X_{n+1,m}^\pm (q)\equiv
\pi q\sum_{j=0}\alpha_j(n,m) (\pm q)^j
.
\label{eq.Xofalpha}
\end{equation}
Only the values with odd $j$ are of interest, because our application
eventually looks only at the differences
$X_{n+1,m}^+(q)-X_{n+1,m}^-(q)$
in which terms of even $j$ cancel.
The basic values for $m=0$ are in Table \ref{tab.alphan0}.

\begin{center}
\begin{table}[hbt]
\begin{tabular}{cc}
$j$ & $\alpha_j(n,0)$ \\
\hline
1 & $(n+1)/4$\\
3 & $n(n+1)(n-2)/32$ \\
5 & $n(n+1)(n-2)^2(n-4)/768$\\
7 & $n(n+1)(n-2)^2(n-4)^2(n-6)/36864$ \\
9 & $n(n+1)(n-2)^2(n-4)^2(n-6)^2(n-8)/2949120$ \\
11 & $n(n+1)(n-2)^2(n-4)^2(n-6)^2(n-8)^2(n-10)/353894400$ \\
\end{tabular}
\caption{
Expansion coefficients of (\ref{eq.Xofalpha}) at $m=0$.
}
\label{tab.alphan0}
\end{table}
\end{center}

Table \ref{tab.alphanm} summarizes (\ref{eq.Xofalpha}) for small values of $n$ and $m$.

\begin{longtable}[l]{p{0.5cm}p{0.5cm}p{16cm}}
$n$ & $m$ & $\sum_{j=1,3,5,7,\ldots}\alpha_j(n,m)q^j$ \\
\hline
0&0&$ +1/4q$\\

0&2&$ +1/4q -1/8q^3$\\

0&4&$ +1/4q -1/2q^3 +1/4q^5$\\

0&6&$ +1/4q -9/8q^3 +3/2q^5 -5/8q^7$\\

0&8&$ +1/4q -2q^3 +5q^5 -5q^7 +7/4q^9$\\

1&0&$ +1/2q -1/16q^3 -1/128q^5 -5/2048q^7 -35/32768q^9 -147/262144q^{11}$\\
&&$ -693/2097152q^{13} -14157/67108864q^{15} -306735/2147483648q^{17} +\ldots$\\

1&2&$ +1/2q -5/16q^3 +7/128q^5 +15/2048q^7 +77/32768q^9 +273/262144q^{11}$\\
&&$ +1155/2097152q^{13} +21879/67108864q^{15} +448305/2147483648q^{17} +\ldots$\\

1&4&$ +1/2q -17/16q^3 +95/128q^5 -245/2048q^7 -483/32768q^9$\\
&&$ -1155/262144q^{11} -3861/2097152q^{13} -62205/67108864q^{15}$\\
&&$ -1130415/2147483648q^{17} +\ldots$\\

1&6&$ +1/2q -37/16q^3 +455/128q^5 -4305/2048q^7 +9933/32768q^9$\\
&&$ +9009/262144q^{11} +20163/2097152q^{13} +255255/67108864q^{15}$\\
&&$ +3926065/2147483648q^{17} +\ldots$\\

1&8&$ +1/2q -65/16q^3 +1407/128q^5 -26565/2048q^7 +213213/32768q^9$\\
&&$ -219219/262144q^{11} -182325/2097152q^{13} -1524237/67108864q^{15}$\\
&&$ -18244655/2147483648q^{17} +\ldots$\\

2&0&$ +3/4q$\\

2&2&$ +3/4q -3/8q^3 +1/8q^5$\\

2&4&$ +3/4q -3/2q^3 +5/4q^5 -3/8q^7$\\

2&6&$ +3/4q -27/8q^3 +45/8q^5 -33/8q^7 +9/8q^9$\\

2&8&$ +3/4q -6q^3 +17q^5 -45/2q^7 +57/4q^9 -7/2q^{11}$\\

3&0&$ +q +3/8q^3 -1/64q^5 -1/1024q^7 -3/16384q^9 -7/131072q^{11}$\\
&&$ -21/1048576q^{13} -297/33554432q^{15} -4719/1073741824q^{17} +\ldots$\\

3&2&$ +q -1/8q^3 +7/64q^5 -13/1024q^7 -19/16384q^9 -35/131072q^{11}$\\
&&$ -93/1048576q^{13} -1221/33554432q^{15} -18447/1073741824q^{17} +\ldots$\\

3&4&$ +q -13/8q^3 +95/64q^5 -625/1024q^7 +1085/16384q^9$\\
&&$ +777/131072q^{11} +1419/1048576q^{13} +15015/33554432q^{15}$\\
&&$ +196625/1073741824q^{17} +\ldots$\\

3&6&$ +q -33/8q^3 +455/64q^5 -6125/1024q^7 +37485/16384q^9$\\
&&$ -30723/131072q^{11} -21021/1048576q^{13} -148005/33554432q^{15}$\\
&&$ -1519375/1073741824q^{17} +\ldots$\\

3&8&$ +q -61/8q^3 +1407/64q^5 -32193/1024q^7 +383229/16384q^9$\\
&&$ -1072071/131072q^{11} +819819/1048576q^{13} +2122263/33554432q^{15}$\\
&&$ +14272401/1073741824q^{17} +\ldots$\\

4&0&$ +5/4q +5/4q^3$\\

4&2&$ +5/4q +5/8q^3$\\

4&4&$ +5/4q -5/4q^3 +5/4q^5 -5/8q^7 +1/8q^9$\\

4&6&$ +5/4q -35/8q^3 +15/2q^5 -55/8q^7 +13/4q^9 -5/8q^{11}$\\

4&8&$ +5/4q -35/4q^3 +25q^5 -75/2q^7 +125/4q^9 -55/4q^{11} +5/2q^{13}$\\

5&0&$ +3/2q +45/16q^3 +45/128q^5 -15/2048q^7 -9/32768q^9 -9/262144q^{11}$\\
&&$ -15/2097152q^{13} -135/67108864q^{15} -1485/2147483648q^{17} +\ldots$\\

5&2&$ +3/2q +33/16q^3 +5/128q^5 +45/2048q^7 -57/32768q^9 -29/262144q^{11}$\\
&&$ -39/2097152q^{13} -315/67108864q^{15} -3245/2147483648q^{17} +\ldots$\\

5&4&$ +3/2q -3/16q^3 +77/128q^5 -735/2048q^7 +3255/32768q^9$\\
&&$ -2009/262144q^{11} -1071/2097152q^{13} -6039/67108864q^{15}$\\
&&$ -50765/2147483648q^{17} +\ldots$\\

5&6&$ +3/2q -63/16q^3 +837/128q^5 -12915/2048q^7 +112455/32768q^9$\\
&&$ -242109/262144q^{11} +147609/2097152q^{13} +312741/67108864q^{15}$\\
&&$ +1756755/2147483648q^{17} +\ldots$\\

5&8&$ +3/2q -147/16q^3 +3245/128q^5 -79695/2048q^7 +1149687/32768q^9$\\
&&$ -4749129/262144q^{11} +9810801/2097152q^{13} -23146695/67108864q^{15}$\\
&&$ -47732685/2147483648q^{17} +\ldots$\\

6&0&$ +7/4q +21/4q^3 +7/4q^5$\\

6&2&$ +7/4q +35/8q^3 +7/8q^5$\\

6&4&$ +7/4q +7/4q^3$\\

6&6&$ +7/4q -21/8q^3 +35/8q^5 -35/8q^7 +21/8q^9 -7/8q^{11} +1/8q^{13}$\\

6&8&$ +7/4q -35/4q^3 +91/4q^5 -35q^7 +133/4q^9 -77/4q^{11} +25/4q^{13}$\\
&&$ -7/8q^{15}$\\

7&0&$ +2q +35/4q^3 +175/32q^5 +175/512q^7 -35/8192q^9 -7/65536q^{11}$\\
&&$ -5/524288q^{13} -25/16777216q^{15} -175/536870912q^{17} +\ldots$\\

7&2&$ +2q +31/4q^3 +119/32q^5 +35/512q^7 +77/8192q^9 -35/65536q^{11}$\\
&&$ -13/524288q^{13} -53/16777216q^{15} -335/536870912q^{17} +\ldots$\\

7&4&$ +2q +19/4q^3 +15/32q^5 +63/512q^7 -483/8192q^9 +777/65536q^{11}$\\
&&$ -357/524288q^{13} -585/16777216q^{15} -2607/536870912q^{17} +\ldots$\\

7&6&$ +2q -1/4q^3 +55/32q^5 -957/512q^7 +9933/8192q^9 -30723/65536q^{11}$\\
&&$ +49203/524288q^{13} -92565/16777216q^{15} -155727/536870912q^{17} +\ldots$\\

7&8&$ +2q -29/4q^3 +559/32q^5 -13585/512q^7 +213213/8192q^9$\\
&&$ -1072071/65536q^{11} +3270267/524288q^{13} -20796633/16777216q^{15}$\\
&&$ +38955345/536870912q^{17} +\ldots$\\

8&0&$ +9/4q +27/2q^3 +27/2q^5 +9/4q^7$\\

8&2&$ +9/4q +99/8q^3 +21/2q^5 +9/8q^7$\\

8&4&$ +9/4q +9q^3 +15/4q^5$\\

8&6&$ +9/4q +27/8q^3$\\

8&8&$ +9/4q -9/2q^3 +21/2q^5 -63/4q^7 +63/4q^9 -21/2q^{11} +9/2q^{13}$\\
&&$ -9/8q^{15} +1/8q^{17}$\\
\caption{
Table of $\sum_j \alpha_j(n,m)q^j$, summed over odd $j$ only.
For odd $n+m$, the series is shown up to $O(q^{17})$, indicated by the triple dots.
}
\label{tab.alphanm}
\end{longtable}

\section{Notations } 
\begin{tabular}{ll}
$\lfloor.\rfloor$ & floor function; largest integer not greater than the argument\\
$(.,.)$           & scalar (inner) product between the two arguments (\ref{eq.Odef})\\
$\mathop{\sum'}_{u=\ldots}$ & summation with the term of index $u=0$ halved\\
$\simeq$          & equivalent upon subtraction of $B^\pm$ or $X^\pm$ of superscripts of opposite sign\\
$A_{n,m}(q)$      & the integral (\ref{eq.Adef}) \\
$\alpha_{n,m}(q)$ & the expansion of the $X$ integral (\ref{eq.Xofalpha}) \\
$B_{n,m}^\pm(q)$      & the integral (\ref{eq.Bdef}) \\
$B(x,y)$& $=\Gamma(x)\Gamma(y)/\Gamma(x+y)$, the Beta Integral \cite[8.38]{GR}\\
$\delta_{jk}$     & Kronecker delta; equal to $1$ if $j=k$, equal to $0$ if $j\ne k$\\
$E(.)$            & complete Elliptic Integral of the first kind \cite[\S 17.3]{AS}\\
${\cal E}$           & electric field amplitude \\
$f_k(q)$          & $k$th orthogonal basis function of the binocular area\\
$F(\begin{array}{cc}a,b\\ c\end{array}\mid .)$          & Gaussian hypergeometric function \cite[\S 15]{AS} \\
$\varphi$	& azimuth angle in spherical coordinates centered at circular sub-pupil \\
$\gamma_{.,.}$    & abscissa sections of the Gram-Schmidt procedure\\
$g_k$             & $k$th input (``seed'') function to the Gram-Schmidt synthesis\\
$g_{.,.}$             & Bessel Function integral \cite[11.3.1]{AS} \\
$I$		& interferometric intensity \\
$i$		& imaginary unit\\
$J_.(.)$		& Bessel function of the First Kind\\
$j!!$             & double factorial, $=1\cdot3\cdot 5\cdot 7 \cdots j$ if $j$ is odd,
 $=2\cdot 4\cdot 6\cdots j$ if $j$ even. $(-1)!!\equiv 1$.  \\
$K(.)$            & complete Elliptic Integral of the second kind \cite[\S 17.3]{AS} \\
$q$               & circle diameter in units of circle center distance\\
$s$		& radial distance to a circle center, $0\le s\le q$\\
$\bm\sigma$, $\sigma_x$, $\sigma_y$, $\sigma$		& wave number, Cartesian coordinates, modulus\\
$T_m(.)$          & Chebyshev Polynomial the first kind of order $m$ \cite[\S 22]{AS} \\
$\theta$	& azimuth angle in the circular coordinates centered at midpoint between sub-pupils\\
$X_{n,m}^\pm(q)$      & the integral (\ref{eq.Xdef}) \\
$Z_k(.)$          & Zernike circle functions \cite{NollJOSA66}\\
$z$		& radial distance to a center of symmetry $0\le z\le 1+q$
\end{tabular}

\bibliographystyle{apsrmp}
\bibliography{all}

\end{document}